\documentclass{article}
\usepackage{appendix}
\usepackage{float}
\usepackage{epsfig,dsfont,amssymb,amsmath,amsfonts,amsbsy,mathrsfs}
\usepackage{graphicx,subfigure}
\usepackage{epstopdf}
\usepackage{mathdots}
\usepackage{float}
\usepackage{quantikz}
\usepackage{hyperref}
\usepackage{cleveref} 
\usepackage{xcolor}
\newtheorem{definition}{Definition}
\newtheorem{theorem}{Theorem}

\usepackage{authblk}
\usepackage{bm}
\usepackage{multirow,booktabs,array}

\usepackage{rotating}
\title{Entangled Mixed-State Datasets Generation by Quantum Machine Learning}
\author[1]{Rui-Bin Xu}
\author[2]{Zheng Zheng}
\author[3]{Yanying Liang  \thanks{Corresponding author: yyl2022@scau.edu.cn}}
\author[1,4]{Zhu-Jun Zheng}
\affil[1]{School of Mathematics, South China University of Technology, Guangzhou, 510641, China.}
\affil[2]{School of Mathematics, Guangdong University of Education, Guangzhou 510303, P.R. China.}
\affil[3]{College of Mathematics and Informatics, South China Agricultural University, Guangzhou 510642, China.}
\affil[4]{Laboratory of Quantum Science and Engineering, South China University of Technology, Guangzhou, 510642, China.}
\date{}

\begin{document}

\maketitle
\begin{abstract}

	The advancement of classical machine learning is inherently linked to the establishment and progression of classical dataset. In quantum machine learning (QML), there is an analogous imperative for the development of quantum entangled datasets comprised with huge quantity and high quality. Especially for multipartite mixed-state datasets, due to the lack of suitable entanglement criteria, previous researchers often could only perform classification tasks on datasets extended based on Werner states or other well-structured states. This paper is dedicated to provide a method for generating mixed-state datasets for entangled-separable classification tasks. This method is based on supervised quantum machine learning and the concentratable entanglement measures. It furthers the assembly of quantum entangled datasets, inspires the discovery of new entanglement criteria with both classical and quantum machine learning, and provides a valuable resource for benchmarking QML models, thereby opening new avenues for exploring the rich structure of quantum entanglement in mixed states. Additionally, we benchmark several machine learning models using this dataset, offering guidance and suggestions for the selection of QML models.
	
\end{abstract}

\section{Introduction}
High-quality large-scale classical datasets, such as the MNIST dataset\cite{deng2012mnist}, ImageNet dataset\cite{deng2009imagenet}, Netflix dataset\cite{bennett2007netflix}, and CIFAR10 dataset, play a crucial role in the extraordinary successes of classical machine learning \cite{atienza2020advanced}. Such available classical datasets also foster novel cross-collaborations among classical machine learning and other disciplines\cite{krizhevsky2012imagenet,cohen2017emnist,li2017cifar10}. 
As a emerging multidisciplinary research area, quantum machine learning promotes the concepts of classical machine learning based on the principles of quantum physics and combines with quantum computing for data analysis in the hope of gaining a potential quantum advantage\cite{biamonte2017quantum,cerezo2022challenges}. Despite the increasing maturity of quantum machine learning, this area still lacks a similarly standardized large-scale datasets for base innovation \cite{perrier2022qdataset}. In addition, the lack of such a large-scale standardized dataset will also hinder opportunities for cross-collaboration between quantum physics, computer science, and other disciplines.

Quantum neural networks (QNN) are deep
learning architectures exhibiting superior learning capabilities\cite{park2024aqua,alrikabi2022face}. Similarly to classical neural networks, QNN is trained by adjusting a given model in a supervised manner until reproducing a given operator. Now, most proposed QNN architectures are benchmarked using classical datasets \cite{jeswal2019recent}. However, the classical training data needs to be encoded into quantum states so that it can be inputted into QNN legally\cite{sun2016quantum,lloyd2020quantum}. Questions have also been raised about whether learning on classical datasets would affect the trainability and advantages of QNN\cite{schatzki2021entangled,kubler2021inductive,mandl2023linear}. 
Meanwhile, some researchers believe that the performance of QNN models can benefit from quantum entangled datasets. When quantum datasets are used for training, Sharma \emph{et al.} pointed out that entanglement in training quantum data can reduce the number of used training datasets\cite{sharma2022trainability}. 
The classic No-free-lunch (NFL) theorem \cite{adam2019no,ouyang2022training,bai2022training} theoretically proves that the performance of any optimizer is related to both the volume of the training data and the degree of match between the training data and the model's inductive bias. This result indicates that starting with quantum entangled datasets may indeed improve the performance of QNN models. In 2024, quantum NFL theorem has been established, which shows the transitional role of quantum entangled datasets in QNN model \cite{wang2024transition}. Specifically, in contrast to previous findings \cite{sharma2022reformulation}, quantum NFL theorem demonstrated that the effect of quantum entangled datasets on prediction error exhibits a double effect, depending critically on the number of measurements allowed. Therefore, the development and supplementation of quantum state datasets may help to unlock the potential advantages of QNN.

Accordingly, some pioneering works have been proposed. The work of Perrier \emph{et al.} in 2022 represents one of the first efforts to generate quantum datasets \cite{perrier2022qdataset}. The authors created 52 datasets derived from simulating single-qubit and two-qubit systems evolved from the Hamiltonian model with or without noise, which can be used for training, benchmarking and competitive development tasks in quantum science. However, the small size of these datasets requires further scaling of qubit numbers. In 2023, Nakayama et al. proposed a quantum circuit classification task, and introduced a smaller quantum circuit datasets with 4, 8, 12, 16, and 20 qubits, which was generated by the famous six Hamiltonian models in condensed matter physics \cite{nakayama2023vqe}. Also in 2023, Placidi \emph{et al.} regarded quantum circuits as unitary operators to introduce a large-scale quantum datasets in the form of quantum assembly language, named MNISQ, which is easy to access \cite{placidi2023mnisq}. However, the field of quantum machine learning still suffers from an absence of comprehensive, large-scale quantum datasets, and such resources remain urgently needed.

We hope to not only generate various quantum entangled datasets, but also reasonably use them to benchmarking different QNN models in the entangled-separable classification task. Related researches on this task based on QML holds great promise. Previous research has largely focused on well-studied quantum states, which restricts the scope and potential applications of their findings. In 2022, Schatzki \emph{et al.} have already taken a step in this direction. They introduce NTangled quantum state datasets that can be used for benchmarking quantum machine learning architectures for supervised learning tasks such as binary classification \cite{schatzki2021entangled}. Despite Schatzki \emph{et al.} established methods for generating pure entangled states, their framework does not address the preparation of entangled mixed states which are more common and worthy of study in the real world. 

In this paper, we provide a complete workflow for the generation of entangled mixed-state datasets. Our approach combines supervised QML and  concentratable entanglement measures. This approach is transferable and scalable, and it offers inspiration for more in-depth classification tasks based on QML. We apply it to entangled-separable classification tasks, where we test three parameterized quantum circuits. We believe that our analysis can provide researchers with recommendations for selecting different models.

In Section.~\ref{PRe}, we introduce concentratable entanglement measures which acts as a special entanglement measures for entangled mixed states. General quantum machine learning framework is also presented here.
Section.~\ref{section_example} mainly illustrates the main results about the analysis about GHZ state and W state with white noise using concentratable entanglement measures, and then promotes this method into generate random entangled states. In Section.~\ref{Numercial}, we give the experimental performance about benchmarking three parameterized quantum circuits on our generated quantum datasets. Section.~\ref{conclude} shows the conclusion and limitations of our work.

\section{Preliminaries}
\label{PRe}
\subsection[1]{Computable Entanglement }
\label{computable:entanglement}
First, we briefly review the concepts of separability and entanglement. Let \( \mathcal{H} = \mathcal{H}_1 \otimes \mathcal{H}_2 \otimes \cdots \otimes \mathcal{H}_n \) be the Hilbert space of an \( n \)-partite quantum system. A quantum state \( \rho \in \mathcal{H} \) is \textbf{fully separable} if it can be expressed as a convex combination of product states:
\begin{equation}\label{separable}
	\rho = \sum_i p_i \, \rho_{1}^{(i)} \otimes \rho_{2}^{(i)} \otimes \cdots \otimes \rho_{n}^{(i)},
\end{equation}
where \( p_i \) is a probability such that \(\sum_i p_i = 1\), and \(\rho_{j}^{(i)}\) are density matrices of subsystem \( j \). Otherwise, \(\rho\) is said to be \textbf{entangled}. It should be noted that, for example, \textbf{\( k \)-separable states (\( k < n \))} also fall within the category of entanglement classification. However, in this paper, we do not make more detailed distinctions and focus solely on the basic classification of separable and entangled states as defined above.

 From the perspective of \textbf{entanglement witnesses}\cite{horodecki2009quantum,branciard2013measurement,sperling2013multipartite,chruscinski2014entanglement,cao2024genuine}, a state \(\rho\) is entangled if there exists a Hermitian operator \(W\) such that \(\text{Tr}(\tau W) \leq 0\) for any separable states \(\tau\), but \(\text{Tr}(\rho W) > 0\). However, finding a suitable entanglement witness \(W\) for arbitrary high-dimensional quantum states is extremely complex and impractical. Consequently, a computable entanglement measure for multipartite pure states was proposed in Ref.~\cite{PhysRevLett.127.140501}.
 
\begin{definition}[\textbf{Concentratable Entanglement}~\cite{PhysRevLett.127.140501}]
	Let $\mathcal{S}$ be the set of qubit indices and $\mathcal{P}(\mathcal{S})$ be the power set of $\mathcal{S}$. For any non-empty subset $s \in \mathcal{P}(\mathcal{S})$, the Concentratable Entanglement of a pure state $\ket{\psi}$ is defined as:
	\begin{equation}
		C_{\ket{\psi}} = 1 - \frac{1}{2^{|s|}} \sum_{\alpha \in \mathcal{P}(\mathcal{S})} \text{tr}[\rho_{\alpha}^{2}],
	\end{equation}
	where $\rho_{\alpha}$ is the reduced state of $\ket{\psi}$ in the subsystems labeled by the elements in $\alpha$, and $|s|$ denotes the number of elements in the set $s$. Note that $\rho_{\emptyset} := 1$.
\end{definition}

\begin{figure}[H]
	\centering
	\begin{quantikz}[row sep={0.6cm,between origins},scale=0.7]\label{CEC} 
		\lstick[4]{Ancillary Register}  & \gate{H} &\ctrl{4}&        &        &\gate{H} &\meter{}\\
		& \gate{H} &        &\ctrl{4}&        &\gate{H}&\meter{}\\
		\vdots \\
		& \gate{H} &        &        &\ctrl{4}&\gate{H} &\meter{}\\
		\lstick[4]{$\ket{\psi}$ } &          &\swap{4}&        &        &&\\
		&          &        &\swap{4}&        &&\\ 
		\vdots \\
		&          &        &        &\swap{4}&&\\ 
		\lstick[4]{ $\ket{\psi}$} &          &\targX{}&        &        &&\\
		&          &		   &\targX{}&        &&\\ 
		\vdots \\
		&          &		   &        &\targX{}&&
	\end{quantikz} 
	\caption{\small \textbf{Parallelized swap test circuit.} The \( n \)-qubit parallelized SWAP test leverages each qubit of ancillary register to conduct a controlled SWAP operation between corresponding qubits of the two \( |\psi\rangle \) copies.}
	\label{fig:CEC}
\end{figure}
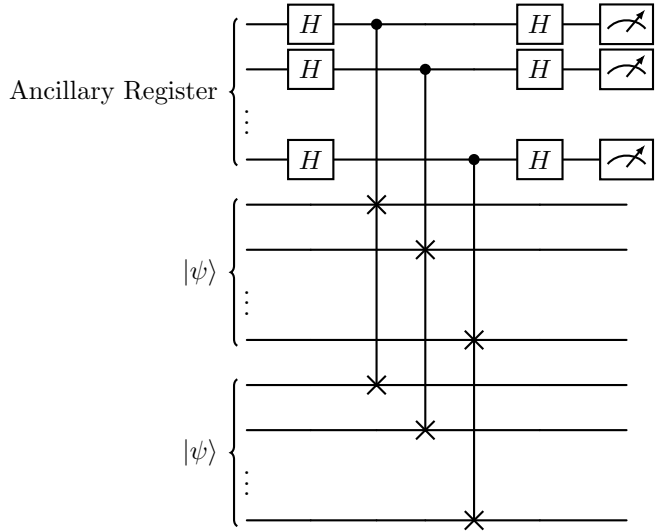

For a given $n$-qubit pure state $\ket{\psi}$, Concentratable Entanglement (CE) measures quantify the average bipartite concurrence\cite{walborn2007experimental,walborn2006experimental} between any possible partition of the whole system, providing an efficient approach to detecting entanglement. Moreover, these measures can be efficiently implemented using a constant-depth circuit as shown in Fig.~\ref{fig:CEC}.

However CE is particularly effective for pure states, the general case for mixed states typically requires a convex roof extension\cite{uhlmann2010roofs}:
\begin{equation}
	C_{\rho}(s) = \inf \sum_{i} p_i C_{\ket{\psi_i}}(s),
\end{equation}
where the infimum is taken over all possible pure-state decompositions of the mixed state $\rho = \sum_{i} p_i \ket{\psi_i}\bra{\psi_i}$. This convex roof construction is often challenging to implement. Instead, several studies \cite{beckey2023multipartite,Foulds_2024} have derived a CE lower bound(CEL) for mixed states, given by:
\begin{equation}\label{lower-bound}
	C^{l}_{\rho} = \frac{1}{2^n} + \left(1 - \frac{1}{2^n} \text{tr}[\rho^2]\right) - \frac{1}{2^n} \sum_{\alpha \in \mathcal{P}(S)} \text{tr}[\rho_{\alpha}^2],
\end{equation}
where $S=n$. Since Eq.~(\ref{lower-bound}) is not an exact bound, it is natural to consider the potential errors associated with its estimation. Therefore, in Sec.~\ref{formulas}, we will investigate two specific types of states to further elucidate this issue.

\subsection{General Quantum Machine Learning Framework}
Now, we present an introduction to Quantum Machine Learning (QML), a methodology that is consistently applied throughout this paper and serves as our means for generating entangled mixed states as well as for classification tasks\cite{schuld2015introduction,biamonte2017quantum,zhang2020recent,batra2021quantum}. QML aims to harness the power of quantum computing to enhance classical machine learning tasks\cite{lloyd2013quantum,schuld2021supervised,alvarez2017supervised}, with the potential to achieve significant speedups for certain problems. A key application of QML is supervised learning, where the goal is to train a Quantum Neural Network (QNN) to classify or predict outputs based on a given dataset of the form \(\{\rho_i, y_i\}\). Here, \(\rho_i \in \mathcal{S} \in \mathcal{H}\) is the quantum state in the corresponding Hilbert space \(\mathcal{H}\), while \(y_i \in \mathcal{Y}\) represents the labels associated with each state according to an map \(f: \mathcal{H} \to \mathcal{Y}\). 

Explicitly, the QNN takes these states \( \rho_i \) in the dataset as input. These states are processed through a unitary transformation \(U(\bm{\theta})\)(referred to as a parametrized quantum circuit or ansatz), where \(\bm{\theta}\) includes continuous parameters that can be optimized during training. The output of the QNN is obtained by measuring a Hermitian observable \(\mathcal{O}\) on the transformed state \(U(\bm{\theta})\rho U^{\dagger}(\bm{\theta})\). For example, the predicted label \(\hat{y}_i\) is often computed using a sign function:
\begin{equation}\label{sig}
	\hat{y}_i = sig\big(\text{Tr}[U(\bm{\theta})\rho_i U^{\dagger}(\bm{\theta})\mathcal{O}]\big),
\end{equation}
which maps the expectation value of the observable \(\mathcal{O}\) to a value between -1 and 1. In improved studies, Eq.~(\ref{sig}) could become more complex, for example by incorporating connections with classical networks to enhance the performance\cite{zhang2023entanglement,lu2018separability}.

To optimize the parameters $\bm{\theta}$ for the classification task over the training set, which is a subset of the whole dataset, a loss function is needed. For binary classification tasks, the loss function is typically defined as the mean-squared error between the predicted and true labels:
\begin{equation}
	L(\bm{\theta}) = \frac{1}{|T|} \sum_{(\rho_i, y_i) \in T} (\hat{y}_i - y_i)^2.
\end{equation}
where \( T \) represents a batch, which is a subset of the training dataset used to compute the loss function \( L(\bm{\theta}) \) at each iteration of the optimization process.
The training process involves minimizing this loss function to optimize the parameters \(\bm{\theta}\). This is achieved by solving the optimization task:
\begin{equation}\label{objective}
	\bm{\theta}^* = \arg \min_{\bm{\theta}} L(\bm{\theta}).
\end{equation}

Quantum Machine Learning serves as a crucial tool in this study. It is utilized in the generation of datasets as well as in the classification tasks of entangled-separable states. Furthermore, in Section.~\ref{Numercial}, we elaborate the performence of different QML models.

\section{Main results about generating entangled mixed states}\label{section_example}
\subsection{Analytical CEL formulas}\label{formulas}
As mentioned in Section.~\ref{computable:entanglement}, we will first present analyses of CEL for Greenberger-Horne-Zeilinger(GHZ) state and W state with white noise, thereby establishing the foundation for our construction of more general entangled mixed states.

\textbf{\textit{GHZ state with white noise}}: 
For a \(n\)-qubit GHZ state with white noise, the density matrix is given by
\begin{equation}
	\rho = p \ket{GHZ_n}\bra{GHZ_n} + \frac{1-p}{2^n} \mathcal{I},
\end{equation}
where \(\ket{GHZ_n} = \frac{1}{\sqrt{2}} (\ket{0}^{\otimes n} + \ket{1}^{\otimes n})\) and \(\mathcal{I}\) is the \(2^n \times 2^n\) identity matrix. It is known that \(\rho\) is fully separable if and only if \(p \leq \frac{1}{1 + 2^{n-1}}\)\cite{dur2000three,pittenger2000note}. We give the lower bound of CE for this state as follows:
\begin{equation}\label{CELGHZformula}
	C^{l}_{\rho} =\frac{p^2}{2} - \frac{1}{2^n} + (1 - p^2) \left( \frac{3}{2^n} - \frac{1}{4^n} - \left( \frac{3}{4} \right)^n \right),
\end{equation}
where \(|S| = n\) is considered. The detailed derivation is provided in Appendix.~\ref{app:details}.
\begin{figure}[H] 
	\centering  
	\vspace{-0.35cm} 
	\subfigtopskip=2pt 
	\subfigbottomskip=2pt 
	\subfigcapskip=-5pt 
	\subfigure[GHZ state with white noise]{
		\label{GHZ.sub.1}
		\includegraphics[width=0.4\linewidth]{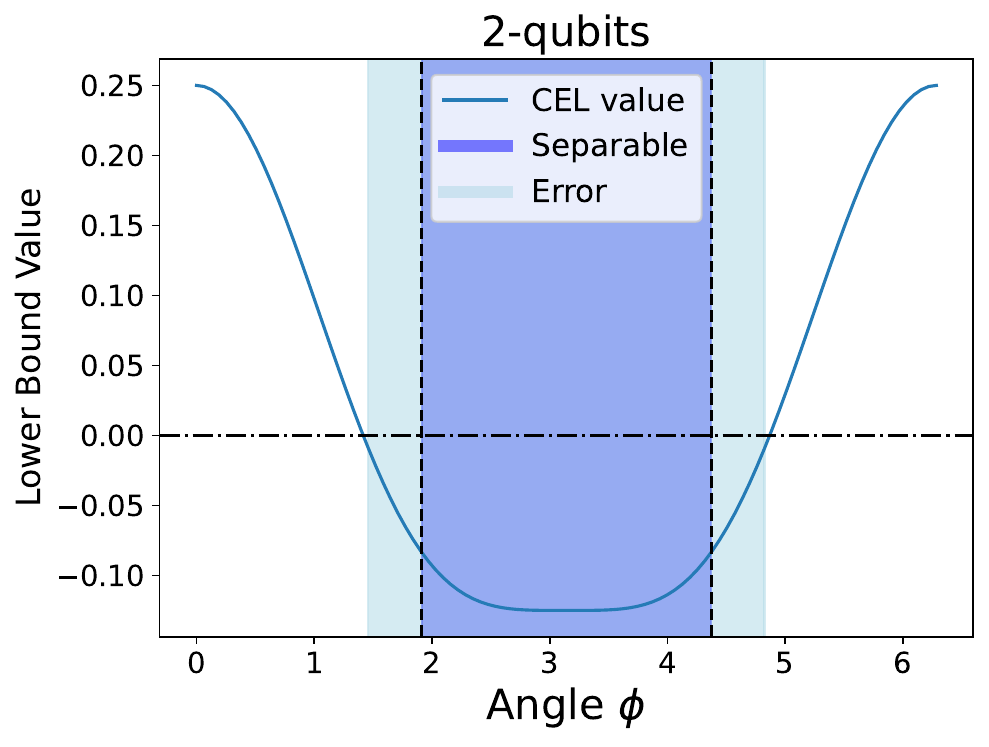}}
	\quad 
	\subfigure[W state with white noise]{
		\label{W.sub.1}
		\includegraphics[width=0.4\linewidth]{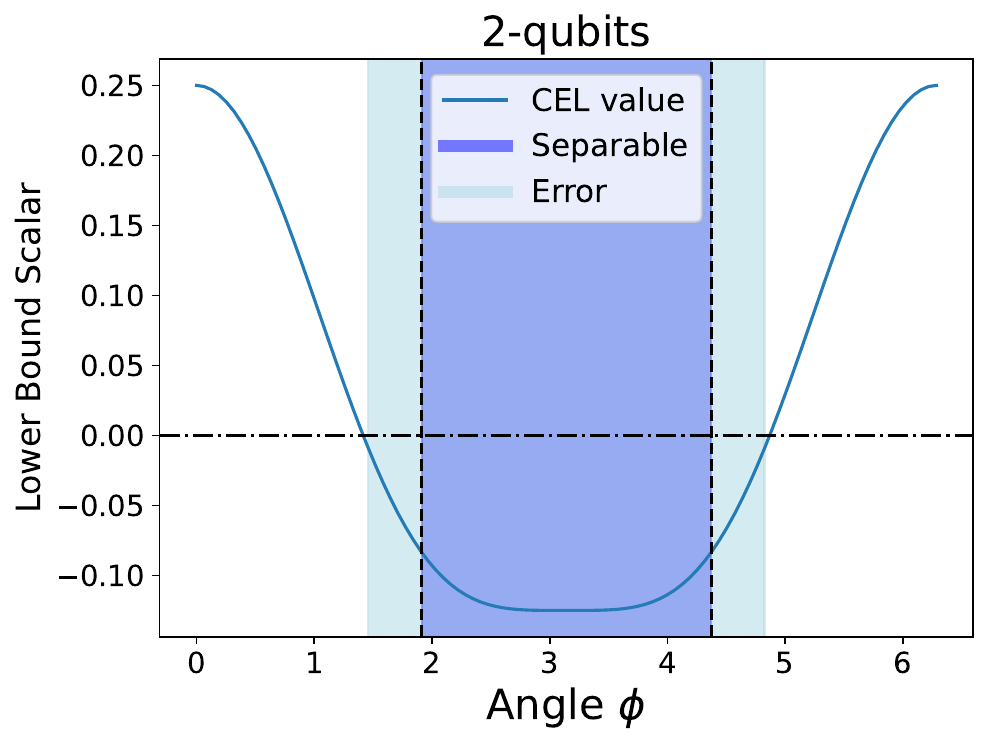}}
	\subfigure[GHZ state with white noise]{
		\label{GHZ.sub.2}
		\includegraphics[width=0.4\linewidth]{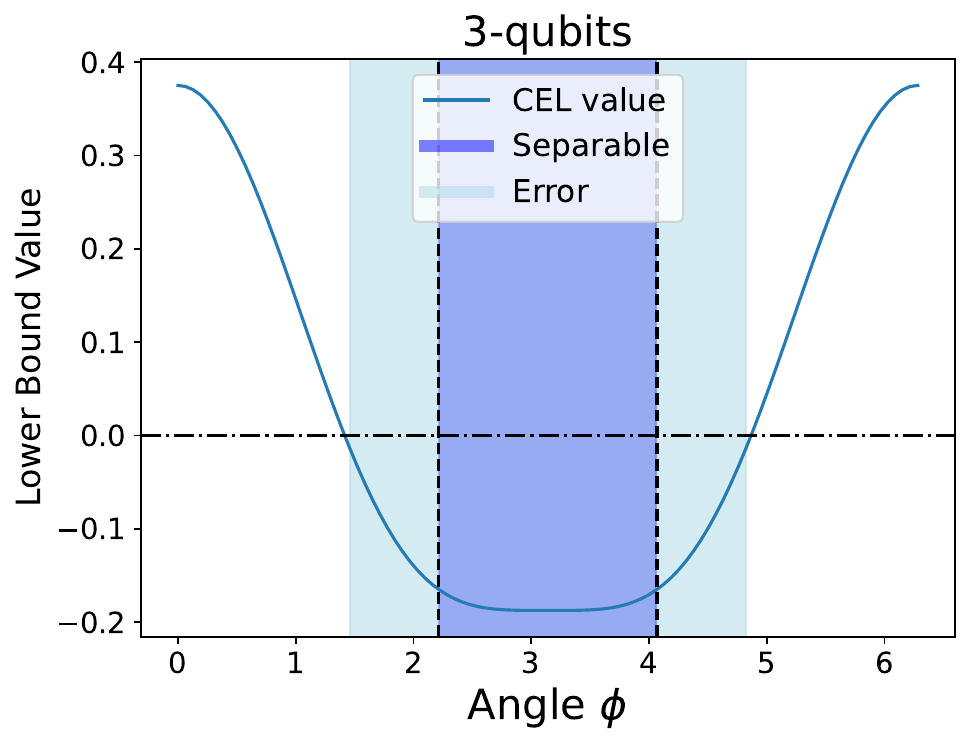}}
	\quad
	\subfigure[W state with white noise]{
		\label{W.sub.2}
		\includegraphics[width=0.4\linewidth]{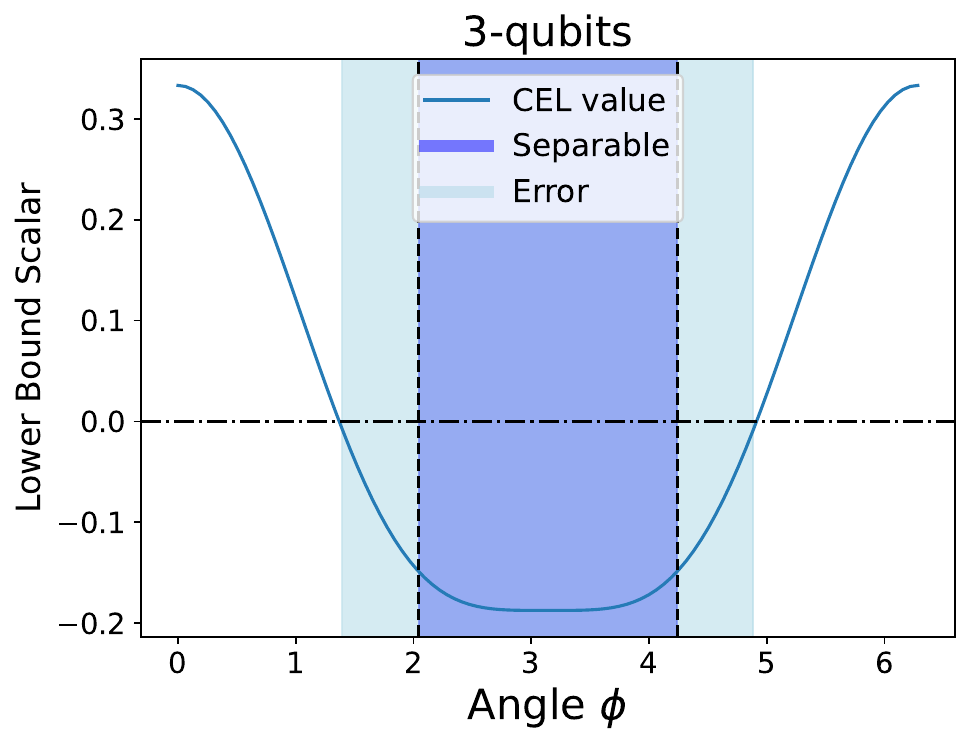}}
	\\
	\subfigure[GHZ state with white noise]{
		\label{GHZ.sub.3}
		\includegraphics[width=0.4\linewidth]{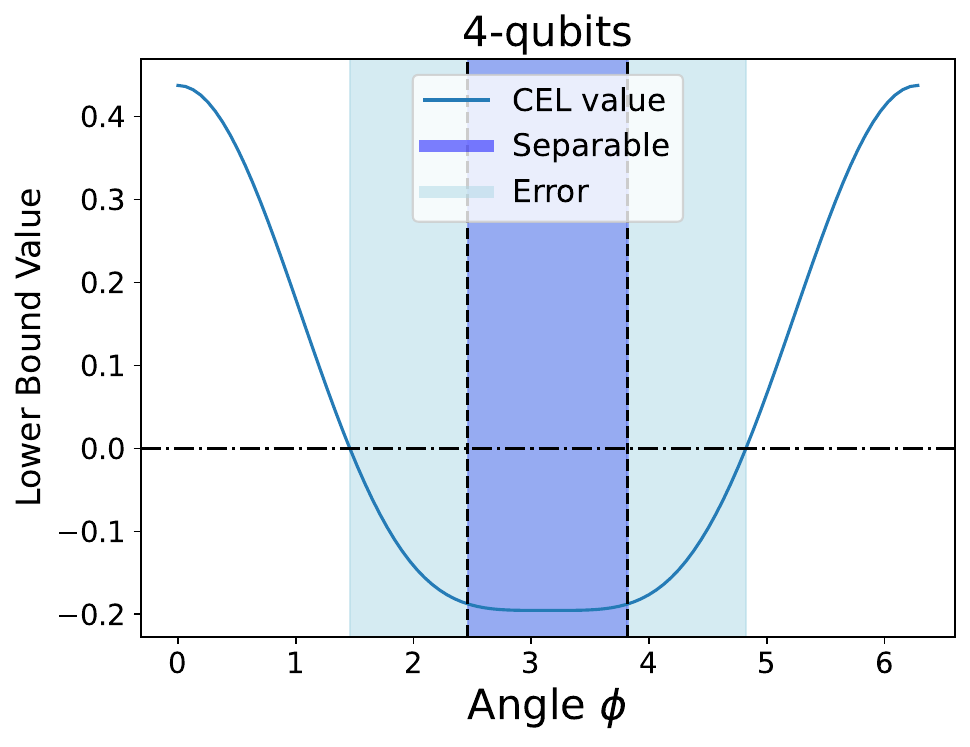}}
	\quad 
	\subfigure[W state with white noise]{
		\label{W.sub.3}
		\includegraphics[width=0.4\linewidth]{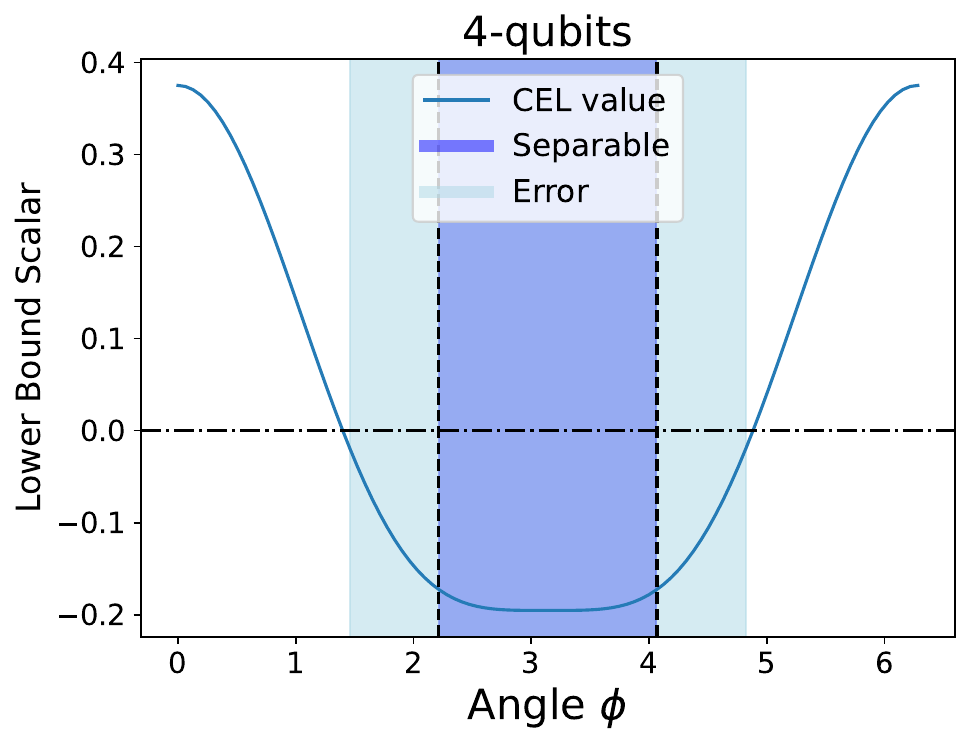}}
	\subfigure[GHZ state with white noise]{
		\label{GHZ.sub.4}
		\includegraphics[width=0.4\linewidth]{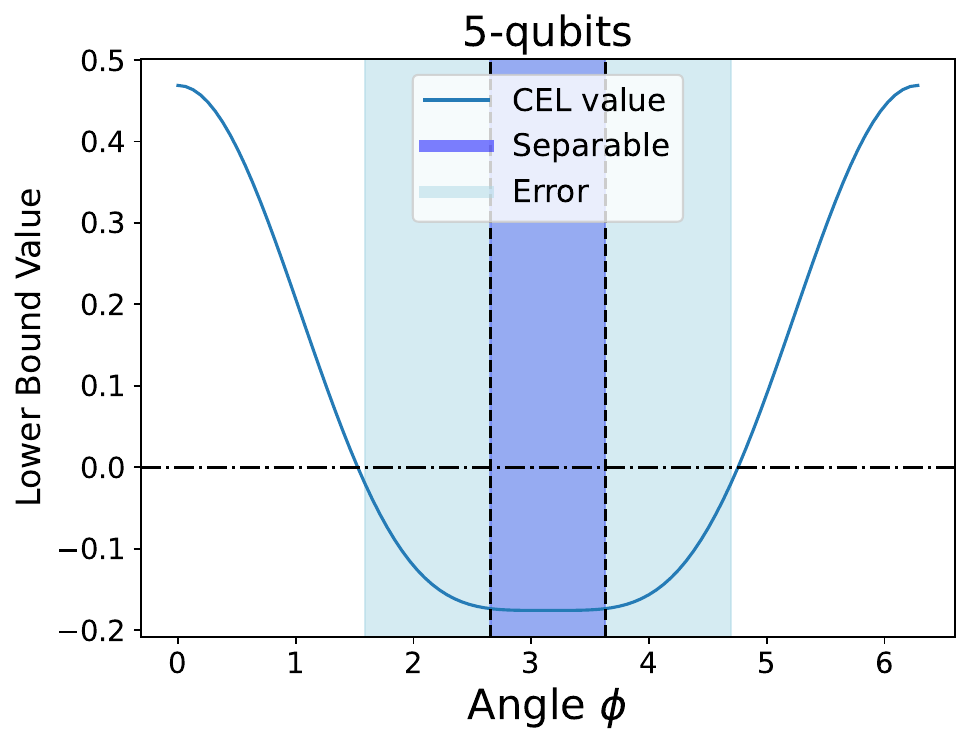}}
	\quad
	\subfigure[W state with white noise]{
		\label{W.sub.4}
		\includegraphics[width=0.4\linewidth]{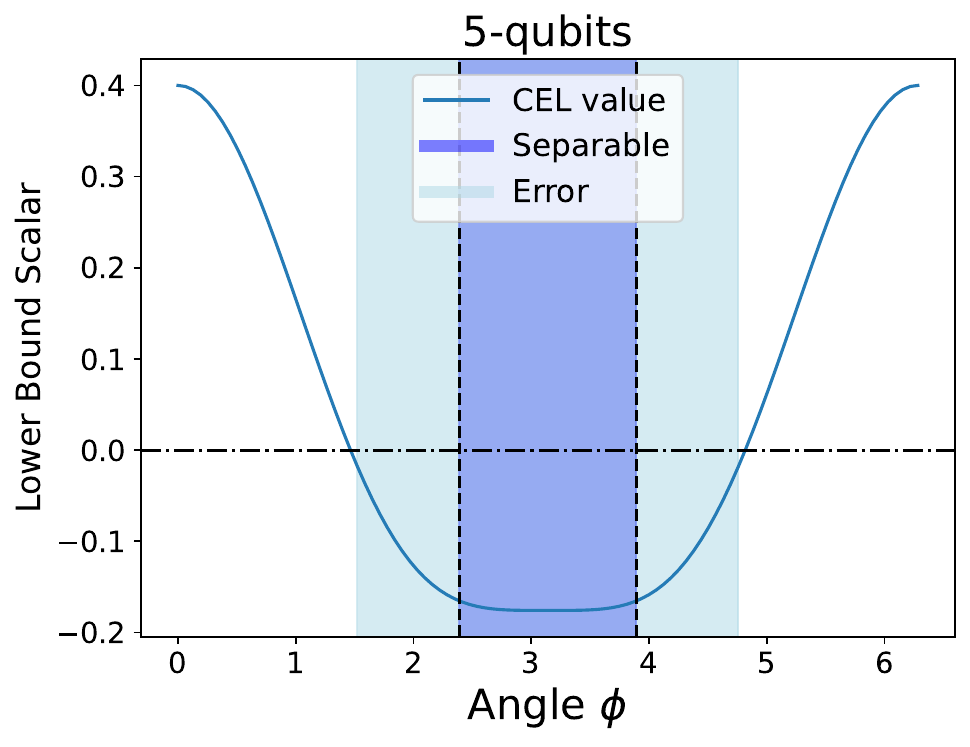}}
	\caption{\small \textbf{CEL for GHZ state and W state with white noise.} This figure shows the numerical simulation of CEL for both types of states for 2, 3, 4, and 5 qubits. $\phi$, where $\phi \in (0, 2\pi)$, represents the parameter of the $R_y$ gate used in the quantum circuit. Correspondingly, the probability $p=cos^{2}(\frac{\phi}{2})$. The blue line represents the CEL value. The blue shaded area represents the actual separable region. The separable region estimated by the CEL, which is larger than the actual one. This error is indicated by the light blue shaded area.}
	\label{level}
\end{figure}

\textbf{\textit{W state with white noise}}: 
For a \(n\)-qubit W state with white noise, the density matrix is given by
\begin{equation}
	\rho = p \ket{W_n}\bra{W_n} + \frac{1-p}{2^n} \mathcal{I},
\end{equation}
where \(\ket{W_n} = \frac{1}{\sqrt{n}} (\ket{100\ldots0} + \ket{010\ldots0} + \ldots + \ket{0\ldots001})\). It is known that \(\rho\) is fully separable if and only if \(p \leq \frac{n}{n + 2^n}\) \cite{gao2010detection}. We give the lower bound of CE for this state as follows:
\begin{equation}\label{CELWformula}
	C^{l}_{\rho} = \frac{(n-1)p^2}{2n} + (1 - p^2) \left( \frac{2}{2^n} - \frac{1}{4^n} - \left( \frac{3}{4} \right)^n \right),
\end{equation}
where \(|S| = n\) is considered. The detailed derivation is provided in Appendix.~\ref{app:details}.

We provide a detailed quantum circuit for a computing CEL for both 2-qubit GHZ states and W states with white noise in Appendix.~\ref{app:details}. The simulation results are in agreement with both Eq.~(\ref{CELGHZformula}) and Eq.~(\ref{CELWformula}). In Fig.~\ref{level}, the maximum values of these states under the quantification of CEL show a significant upward trend with increasing system size, as observed in each column. Meanwhile, the error of CEL are becoming more evident. For 5-qubit case, as shown in Fig.~\ref{GHZ.sub.4} and Fig.~\ref{W.sub.4}, nearly half of the separable states are identified as entangled states, if we use CEL as a criterion. However, the method used to identify entangled states is absolutely accurate. Therefore, we could combine CEL and supervising QML as a approach for generating entangled mixed-state datasets.

\subsection{Generating random entangled mixed states using quantum circuit}\label{method}
In this section, associated with factors such as generation efficiency, circuit depth, and decoherence associated with multiple quantum gates in practical applications, we propose an efficient method for generating a large number of entangled mixed states with desired distribution quantified by CEL within quantum circuits.

According to the purification theorem in Ref.~\cite{Nielsen_Chuang_2010}, any arbitrary state $\rho_T$ in a finite-dimensional Hilbert space $H_T$ can always be purified into a pure state $|\Psi_{TA}\rangle$ in an enlarged Hilbert space $H_{TA} \cong H_T \otimes H_A$, such that $\rho_T$ is a partial trace of $|\Psi_{TA}\rangle$, i.e., $\rho_T = \text{Tr}_A |\Psi_{TA}\rangle \langle\Psi_{TA}|$. From the perspective of designing quantum circuits, \( H_T \) and \( H_A \) can be regarded as the \textbf{target register} and the \textbf{ancillary register}, respectively. Conversely, a mixed state can be derived from pure states in the circuit through the process of reducing the ancillary register. To some extent, the method of obtaining mixed states in a quantum circuit is actually a form of \textit{anti-purification}. This idea is also embodied in the design of the two mixed-state generation circuits described in Section~\ref{formulas} and other related research\cite{riedel2021bell,cruz2019efficient}.
The entanglement between the ancillary register which will be reduced and the target register plays an important role in our design for generating mixed states in quantum circuit.
This motivates us to generate mixed states from entangled pure states derived by different QNNs.

\begin{figure}[H] 
	\centering  
	\vspace{-0.35cm} 
	\subfigtopskip=2pt 
	\subfigbottomskip=2pt 
	\subfigcapskip=-5pt 
	\subfigure[Hardware efficient ansatz]{
		\label{fig:hardware_efficient}
		\includegraphics[width=0.3\linewidth,height=3cm]{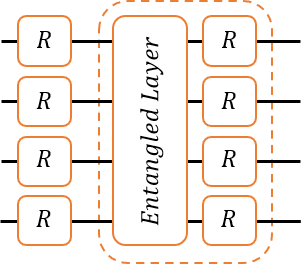}}
	\quad 
	\subfigure[Strongly-entangling ansatz]{
		\label{fig:strongly_entangling}
		\includegraphics[width=0.5\linewidth,height=3cm]{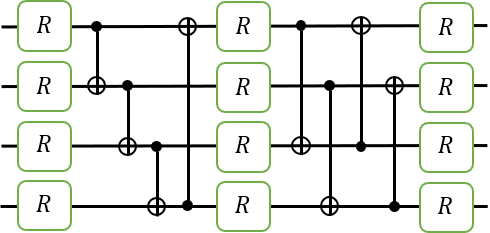}}
	\subfigure[Simplified 2-design ansatz]{
		\label{fig:simplified_2_design}
		\includegraphics[width=0.6\linewidth,,height=3.9cm]{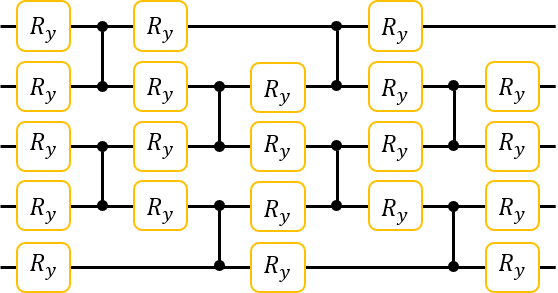}}
	\quad
	\caption{\small \textbf{Three different ans\"atz for mixed-states generation}}
	\label{fig:ansatz}
\end{figure}

We tested the performance of three different ans\"atze in generating mixed states with variable dimension at different numbers of layers:

1. \textbf{Hardware-efficient ansatz (HWE)\cite{kandala2017hardware} in Fig.~\ref{fig:hardware_efficient}}: Composed of layers of single-qubit rotations denoted by $R$ and entangled layers consisted with two-qubit entangling gates such as CNOTs, it is designed to be efficient in terms of circuit depth and gate count, making it suitable for near-term quantum hardware.

2. \textbf{Strongly-entangling ansatz (SEA)\cite{schuld2020circuit} in Fig.~\ref{fig:strongly_entangling}}: Composed of arbitrary single-qubit rotations denoted by $R$ and varied pattern CNOT entanglement, it is designed to maximize entanglement between qubits and generate highly entangled states.

3. \textbf{Simplified 2-design ansatz (SD)\cite{cerezo2021cost} in Fig.~\ref{fig:simplified_2_design}}:Composed of Pauli-Y rotations denoted by $R_y$ and entanglers bwtween neighbor qubits, it is commonly used to study barren plateaus in quantum optimization. 

\begin{figure}[H]
	\centering
	\includegraphics[width=1\linewidth]{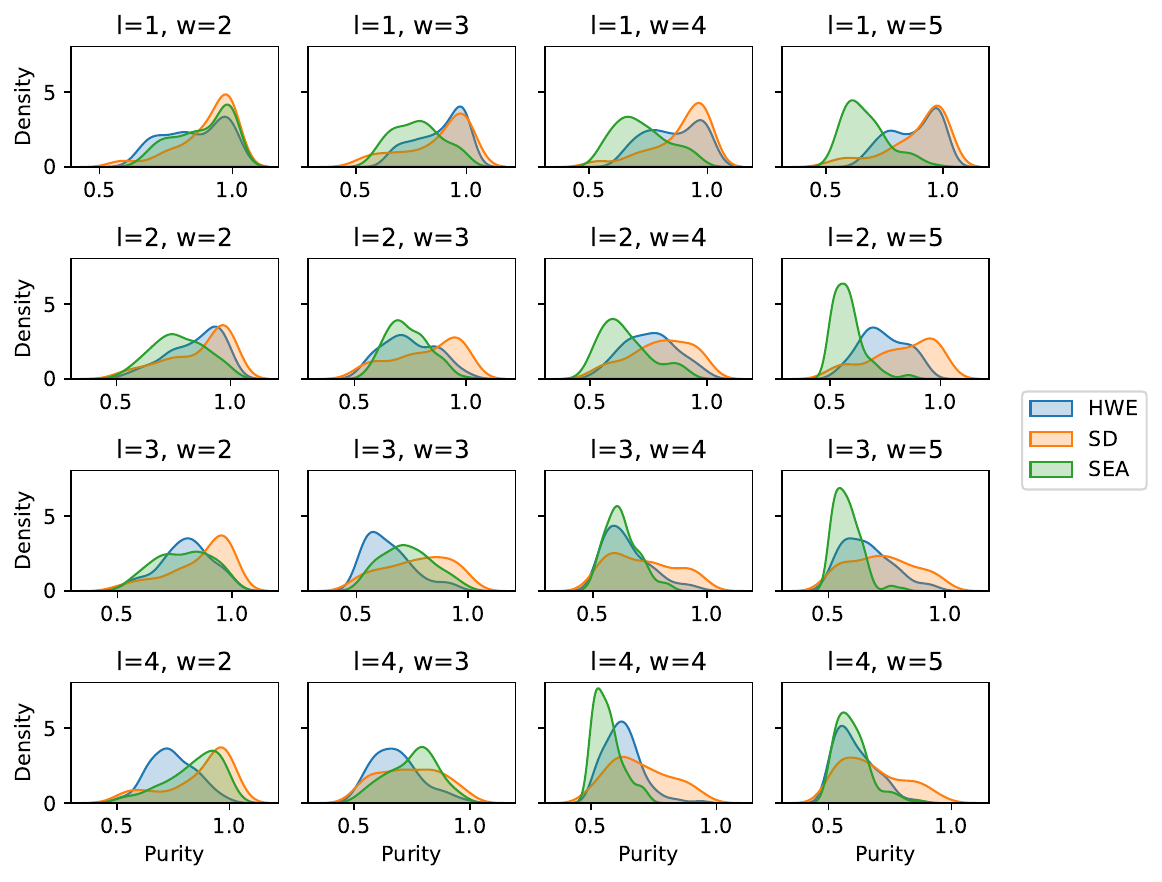}
	\caption{\small \textbf{Purity distribution of randomly generated mixed states.} These states are generated by SEA, HWE, and SD ansatz with different depth (1, 2, 3, 4) and width (2, 3, 4, 5). The blue, green, and orange regions represent the kernel density estimates for HWE, SEA, and SD, respectively. Each group contains 100 quantum states.}
	\label{fig:comparediffrentansatz1}
\end{figure}

In this simulation, initial state is $\ket{0}\bra{0}$. The first qubit will be set as auxiliary register, for any mentioned ans\"atze. For each type, we considered different depth (denoted by \(l\), with \(l = 1, 2, 3, 4\)) and width (denoted by \(w\), with \(w = 2, 3, 4, 5\)). For all circuits, we randomly generated 100 states, and the purity distributions of them are illustrated in Fig.~\ref{fig:comparediffrentansatz1}. The SEA(the green area) is notably susceptible to variations in both depth and width. As both depth and width increase, the distribution of purity for the SD(the orange area) shifts towards lower values. The performance of the SD ansatz is less affected by varied width, but more affected by depth. The HWE(the blue area) , however, remains largely invariant with respect to variations in width. Overall, the three types of ansatz are all affected by depth and width, yet they are all capable of generating mixed states.

Before delving into the specific methods, we firstly need to establish the theoretical foundation supporting our approach. We derive the following continuity of CEL:

\begin{theorem}\label{thm}
	Given two n-qubit state $\rho$ , $\sigma$, if $D_{tr}(\rho,\sigma)\leq d$, then $|C^{l}_{\rho}(S)-C^{l}_{\sigma}(S)|\leq (\frac{1}{2^n}+1)\sqrt{2}d$, where $D_{tr}(\cdot)$ denotes the trace distance.
\end{theorem}
 The detailed derivation is provided in Appendix.~\ref{app:Thm}. Therefore, based on the unitary invariance, we can conclude that when the initial states are sufficiently close, the CELs of two quantum states after the action of a unitary operator are also close. So, we could firstly generate an mixed state $\rho_{\text{in}}$ with an desired CEL value. By applying different perturbations $\epsilon$ to the initial state through local transformations, we can obtain an initial state dataset $P$, making it satisfies: $\mathop{max}_{\rho' \in P}Dr(\rho^{'},\rho_{in})\leq d$. In this way, to generate a mixed-states dateset with target CEL value $\xi$, we just need to train the parameters such that $|C_{\sigma}^{l}(S) -\xi| \leq \delta$, where $\sigma = U(\bm{\theta})\rho_{in} U^{\dagger}(\bm{\theta})$ and $\delta$ is the tolerant value.

\begin{figure}[htpb]
	\centering
	\includegraphics[width=0.8\linewidth]{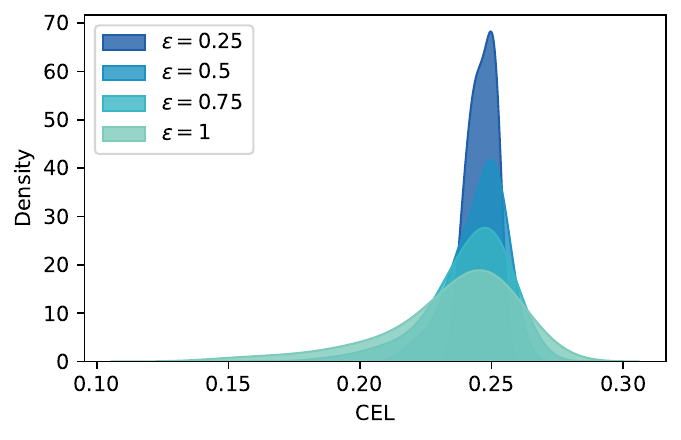}
	\caption{\small \textbf{Comparison of different $\epsilon$ values on state distribution.} This figure illustrates the performance of the SEA in generating specific distributions of entangled mixed states under different perturbations $\epsilon$. The initial state with a desired CEL value as 0.25. the local unitary operations are single-qubit rotations and $\epsilon$ are chosen from $\{0.25, 0.5, 0.75, 1\}$. Each group of data consists of 1000 states, distributed around $\xi = 0.25$. }
	\label{disturb}
\end{figure}

For the optimized parameterized quantum circuit, we can conclude that any set of close initial states, after being acted upon by this parameterized quantum circuit, will be distributed around $\xi$. Therefore, using this method, we can train the parameterized quantum circuit for a target entangled mixed-state dataset. In Fig.~\ref{disturb}, we show the performance of the SEA in generating 3-qubit entangled mixed states. For simplicity, we did not strictly calculate the trace distance between the initial states when generating the initial state set. Instead, we controlled the angles of the rotation operations applied to the initial states, with the rotation angles randomly selected from the interval $(-\epsilon, \epsilon)$. As expected, as $\epsilon$ increases, the distribution of the set composed of 1000 states becomes more and more broader. This helps us select an appropriate degree of perturbation.

Based on the aforementioned methods, we can generate a large number of different mixed-state entangled datasets. Depending on different requirements, in the future one can modify our strategies on this basis to enhance their practicality.

\section{Main results about benchmarking QML model using generated entangled mixed states}\label{Numercial}
In this section, we conduct a entanglement-separability classification benchmark test on the entangled mixed-state datasets we have generated, which include four datasets of 2, 3, 4, and 5 qubits. In the generation of datasets, the first two qubtits are set as ancillary register and initial state is $\ket{0}$. The composition of our dataset is as follows:

\textbf{Separable mixed-state dataset}: All separable states in our dataset are generated based on Eq.~(\ref{separable}). To avoid entanglement, only random controlled rotation gates are implemented between ancillary register and target register. Specifically, the control qubit is randomly chosen from the ancillary register, and the target qubit from the target register. For every qubit system, the dataset contains 6000 states. 

\textbf{Entangled mixed-state dataset}: We set $\epsilon = 0.5$ and $\xi = 0.25$ in construction entangled mixed-state datasets. For each qubit dataset, it contains a total of 6,000 quantum states generated by SD, HWE, and SEA, respectively, with different numbers of layers (see Table \ref{tab:quantum_states} about five qubit dataset). These states are designed to have varying degrees of entanglement quantified by the CEL.
\begin{table}[htbp]
	\centering
	\caption{Dataset Structure}
	\label{tab:quantum_states}
	\begin{tabular}{ccccc}
		\toprule
		\textbf{Dataset} & \textbf{Ansatz Type} & \textbf{Width} & \textbf{Depth} & \textbf{Count} \\
		\midrule
		\multirow{12}{*}{FIVEQUBITDATA} & \multirow{4}{*}{HWE} & \multirow{4}{*}{7} & 2 & 500 \\
		& & & 3 & 500 \\
		& & & 4 & 500 \\
		& & & 5 & 500 \\
		\cmidrule{2-5}
		& \multirow{4}{*}{SD} & \multirow{4}{*}{7} & 2 & 500 \\
		& & & 3 & 500 \\
		& & & 4 & 500 \\
		& & & 5 & 500 \\
		\cmidrule{2-5}
		& \multirow{4}{*}{SEA} & \multirow{4}{*}{7} & 2 & 500 \\
		& & & 3 & 500 \\
		& & & 4 & 500 \\
		& & & 5 & 500 \\
		\bottomrule
	\end{tabular}
\end{table}

In classification, we choose a simple observable $\mathcal{O} = Z$, where $Z$ is the Pauli operator acting on the last qubit of target register, leaving the rest Identity, i.e. $Z=I\otimes I\otimes\dots\otimes \sigma_z$. This choice of observable is motivated by its simplicity and the ease of measurement in experimental setups.
\begin{figure}[H]
	\centering
	\vspace{-0.35cm}
	\subfigtopskip=2pt
	\subfigbottomskip=2pt
	\subfigcapskip=-5pt
	
	\subfigure[2-qubit dataset]{
		\label{con1}
		\includegraphics[width=0.4\linewidth]{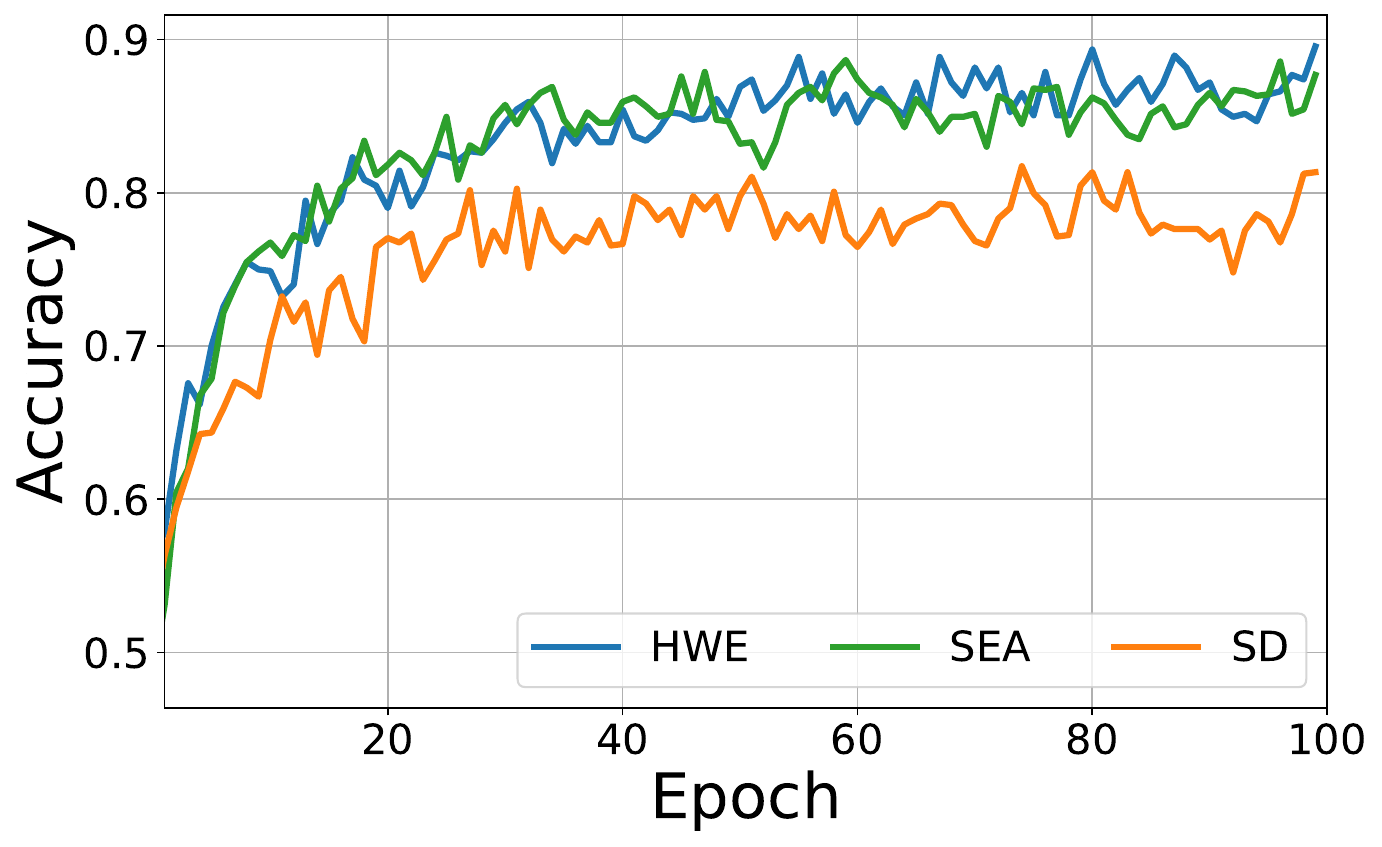}}
	\quad
	\subfigure[3-qubit dataset]{
		\label{con2}
		\includegraphics[width=0.4\linewidth]{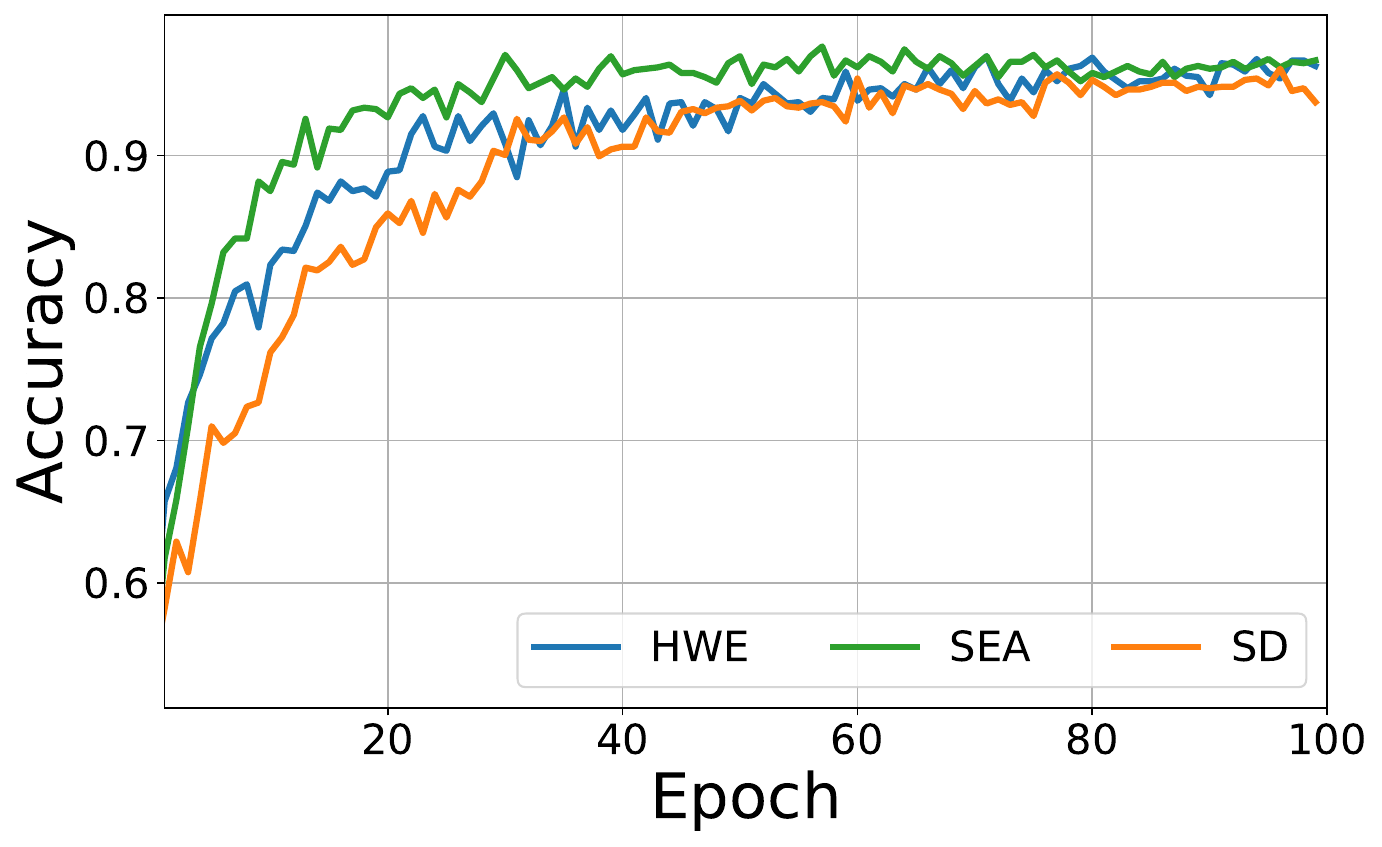}}
	
	\subfigure[4-qubit dataset]{
		\label{con3}
		\includegraphics[width=0.4\linewidth]{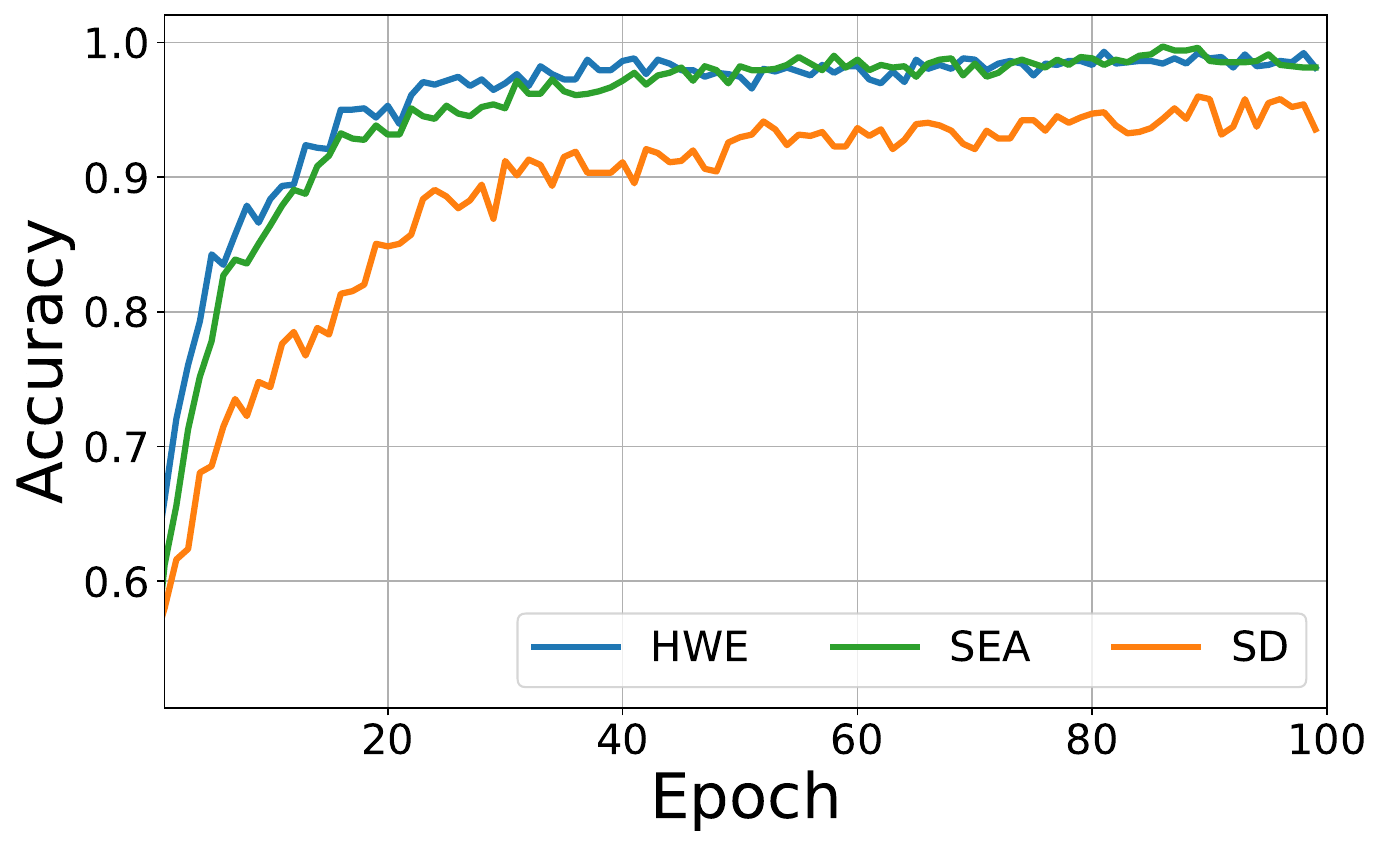}}
	\quad
	\subfigure[5-qubit dataset]{
		\label{con4}
		\includegraphics[width=0.4\linewidth]{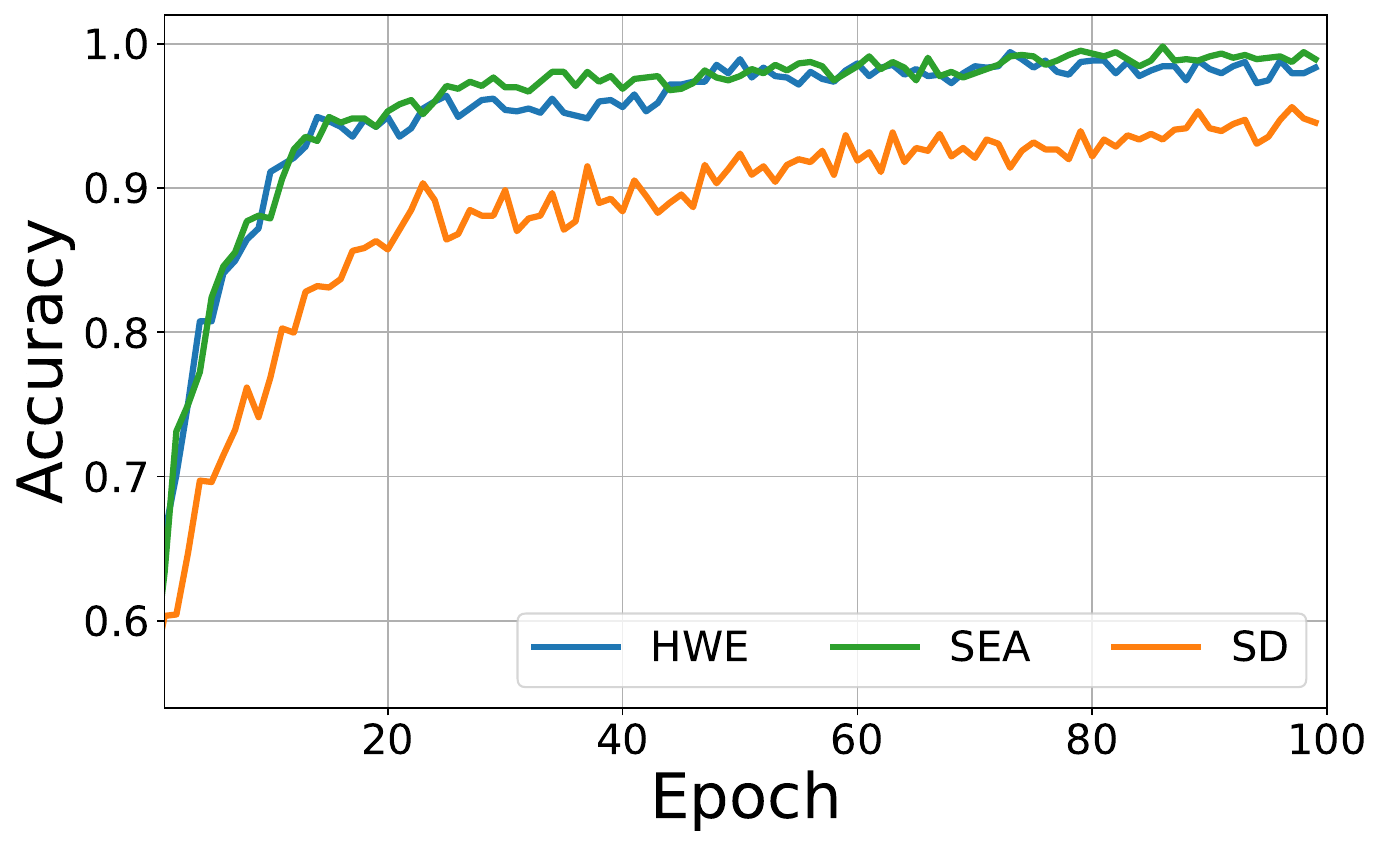}}
	
	\caption{\small \textbf{Convergence of three different ans\"atz.} The figure presents the evolution of average batch accuracy with respect to the number of 100 iterations for three different ans\"atz during training across various datasets.  The blue line represents the performance of the HWE, the green line represents the performance of the SEA, and the orange line represents the performance of the SD. Figs.~\ref{con1} to \ref{con4} show the performance on datasets of 2, 3, 4, and 5 qubits, respectively.}
	\label{convergence}
\end{figure}
\textbf{Convergence Behavior:} First, we analyzed the convergence behavior of three types of ansatz on different qubit datasets.
To evaluate the overall performance of different types ansatz, we calculated the average success rate of each circuit on each training batch and showed how these rates changed with the number of iterations.
Specifically, for the \(n\)-qubit datasets, where \(n=2,3,4,5\), the range of circuit widths is $\{n,n+1,n+2,n+3\}\}$, while the depth is independent of \(n\) and the range of depths is \(\{3, 4, 5, 6\}\). These average success rates were computed across all training batches for different combinations of widths and depths for each type of ansatz, with a batch size of 32 used in our experiments, as illustrated in Fig.~\ref{convergence}. Within the first 20 iterations, the accuracy increased rapidly. As the number of iterations increased, the accuracy gradually stabilized, with the final stable success rate increasing with the system size. This trend is evident from the stable accuracy of each ansatz depicted in Figs.~\ref{con1} to \ref{con4}. This is primarily because the width of the circuits varies with the system size when setting the training hyperparameters. This variation in width directly impacts the training efficiency of the model. Overall, in training, the training efficiency of SD(the orange line) is relatively low, especially in Figs.~\ref{con1}, \ref{con3} and \ref{con4}, while the training efficiencies of HWE and SEA are comparable.

\textbf{Architecture Analysis:} To analyze the impact of different architectures on the performance of three types of ansatz, we specifically varied the depths and widths of these models. We used the accuracy and \textbf{F1 Score}\cite{vakili2020performance,sokolova2006beyond} of the trained models on the test set as evaluation criteria. The F1 Score is a metric that combines precision and recall, commonly used in classification tasks. Its mathematical form is:
\[
\text{F1 Score} = 2 \times \frac{\text{Precision} \times \text{Recall}}{\text{Precision} + \text{Recall}}
\]
where Precision is \(\frac{\text{TP}}{\text{TP} + \text{FP}}\) and Recall is \(\frac{\text{TP}}{\text{TP} + \text{FN}}\). Here, TP, FP, and FN represent true positives, false positives, and false negatives, respectively. While our dataset is balanced, relying solely on accuracy can still be insufficient for a comprehensive evaluation of model performance, as it does not account for the balance between precision and recall, which is crucial for understanding the model's ability to correctly identify positive instances without excessive false positives or false negatives. 

\begin{figure}[] 
	\centering  
	\vspace{-0.35cm} 
	\subfigtopskip=2pt 
	\subfigbottomskip=2pt 
	\subfigcapskip=-5pt 
	\subfigure[]{
		\label{Width:2}
		\includegraphics[width=0.43\linewidth]{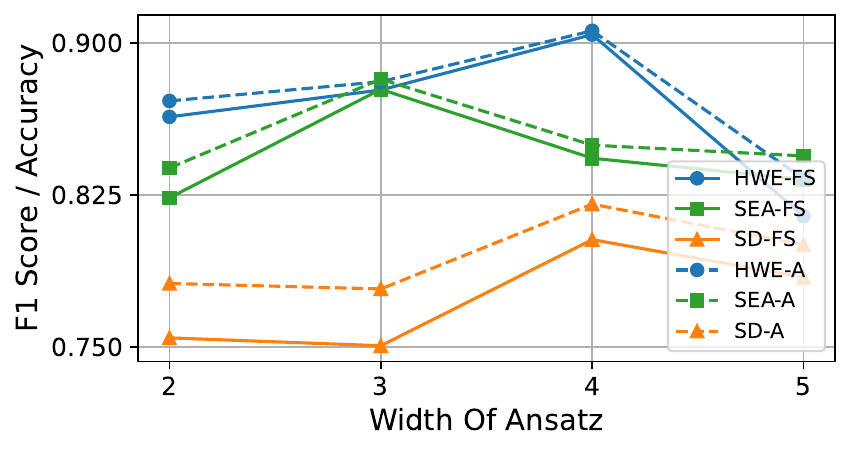}}
	\quad 
	\subfigure[]{
		\label{Depth:2}
		\includegraphics[width=0.43\linewidth]{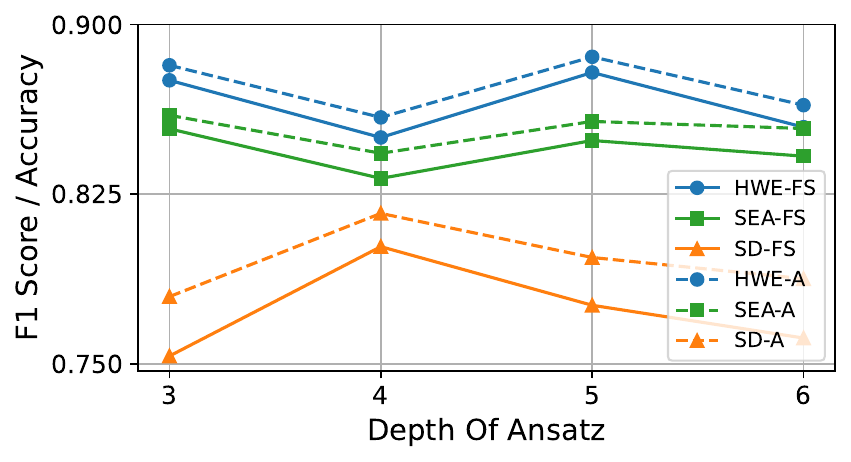}}
	\subfigure[]{
		\label{Width:3}
		\includegraphics[width=0.43\linewidth]{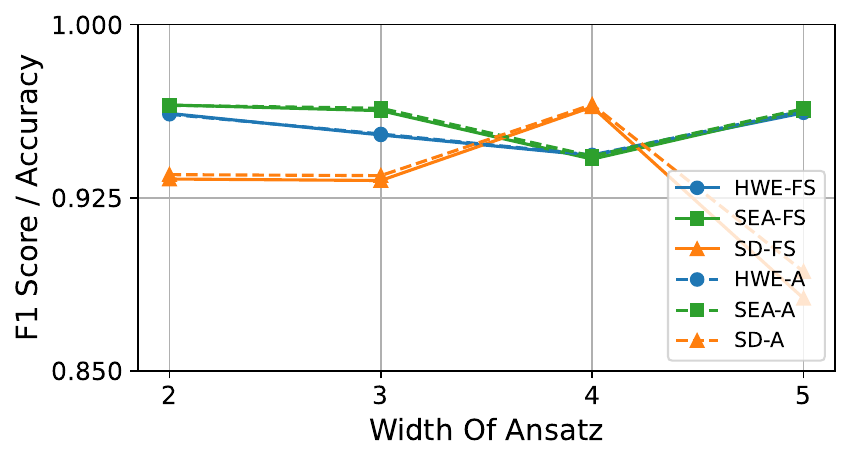}}
	\quad
	\subfigure[]{
		\label{Depth:3}
		\includegraphics[width=0.43\linewidth]{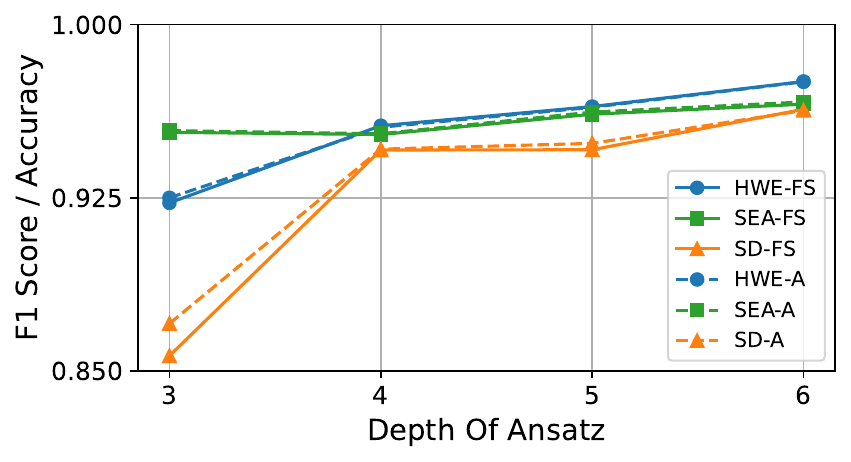}}
	\\
	\subfigure[]{
		\label{Width:4}
		\includegraphics[width=0.43\linewidth]{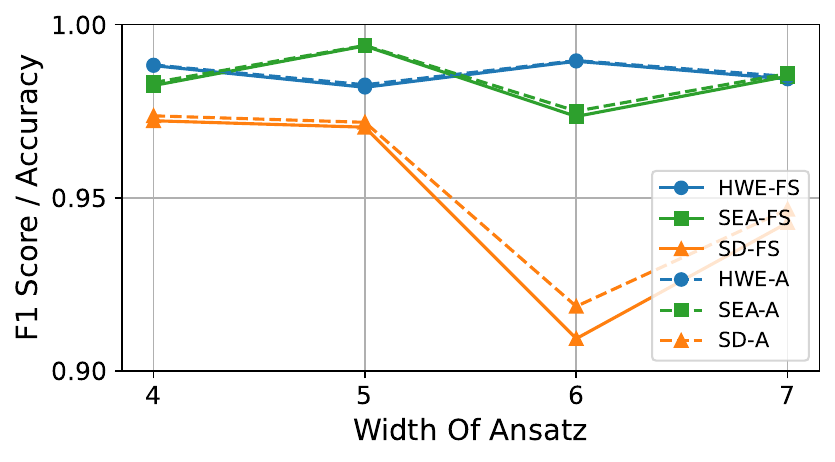}}
	\quad 
	\subfigure[]{
		\label{Depth:4}
		\includegraphics[width=0.43\linewidth]{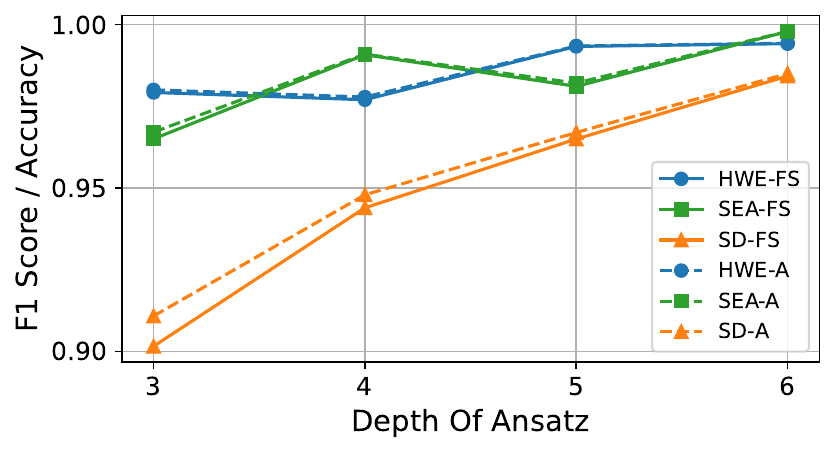}}
	\subfigure[]{
		\label{Width:5}
		\includegraphics[width=0.43\linewidth]{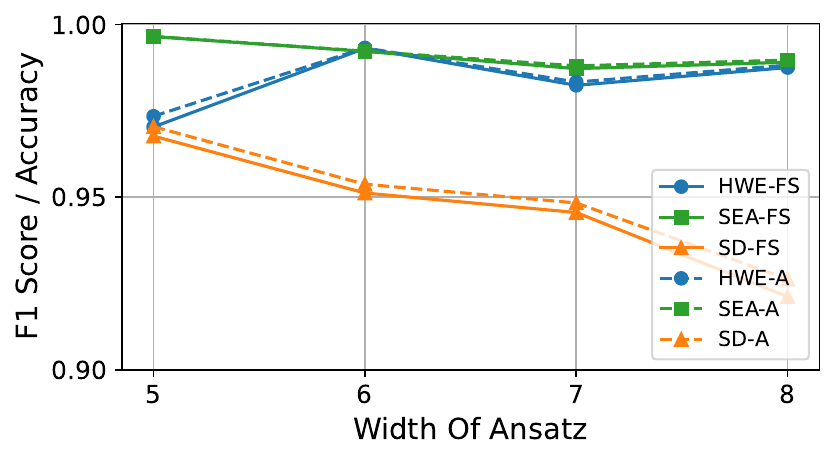}}
	\quad
	\subfigure[]{
		\label{Depth:5}
		\includegraphics[width=0.4\linewidth]{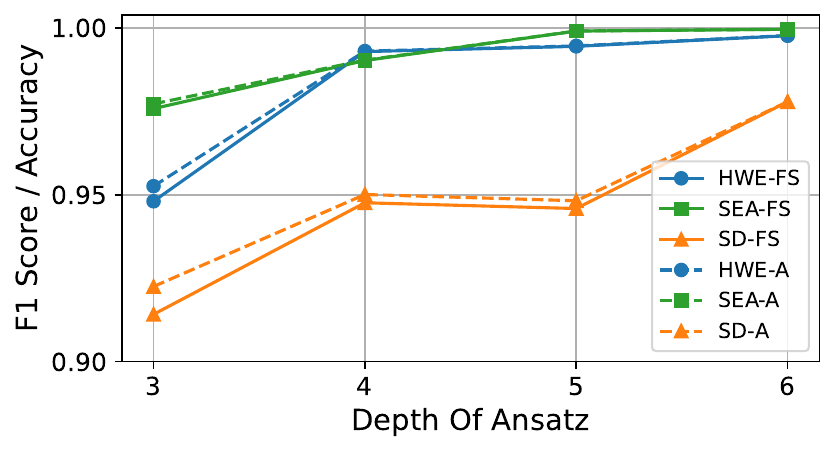}}
	\caption{\small \textbf{Performance of varied width and depth.} This figure shows the accuracy and F1 score of three ans\"atz on the test set as the width and depth vary. The blue lines represents the performance of the HWE, the green lines represents the performance of the SEA, and the orange lines represents the performance of the SD. Solid lines represent F1 scores, while dashed lines represent accuracies.}
	\label{convergence1}
\end{figure}

In Fig.~\ref{convergence1}, we observe that for different widths, the accuracy and F1 Score of the three models do not show significant improvements. Meanwhile, as the system size increases, the accuracy and F1 Score of the SD model exhibit large fluctuations, In Figs.~\ref{Width:4} and \ref{Width:5}, as the width increases, the accuracy and F1 Score of the SD(the orange) even decrease. While the SEA and the HWE demonstrate more stable performance. Variations in depth have a more pronounced impact on the accuracy and F1 Score of the model, with deeper models achieving higher accuracy and F1 Score as the system size increases.  This trend is particularly evident in Fig.~\ref{Depth:4} and Fig.~\ref{Depth:5}, where the impact of depth variations on model performance is clearly illustrated. In contrast, the SD model not only has lower convergence but also its accuracy is significantly affected by both depth and width. Therefore, we conclude that in this work, the depth of the ansatz plays a primary role in its learning performance. 

Overall, through our numerical simulations, we find that HWE and SEA perform better in the dataset classification task, with their performance mainly influenced by depth. In contrast, the SD model not only has lower convergence but also its accuracy is significantly affected by both depth and width, and is more sensitive to changes in different architectures. Using SD for related work may increase additional debugging workload.

Additionally, we also conducted learning on both GHZ state and W state with white noise. However, the highest accuracy in this architecture is only $60\%$, which is equivalent to random guessing in most cases. This phenomenon is not surprising, and we provide the following explanation: 

In previous studies, a ansatz denoted by $U(\bm{\theta})$, is generally considered to act on the quantum state $\rho$. However, based on the cyclic property of the trace operation, the measurement outcome $\text{Tr}(U(\bm{\theta})\rho_i U^{\dagger}(\bm{\theta})\mathcal{O})$ can be equivalently expressed as $\text{Tr}(\rho \mathcal{O(\theta)})$, where $\mathcal{O}(\theta)=U^{\dagger}(\theta)\mathcal{O}U(\theta)$. This implies that $U(\bm{\theta})$ can also be interpreted as acting on the observable $\mathcal{O}$. For quantum states of two different classes, the goal of QML is to optimize the parameters $\bm{\theta}$ such that $\text{Tr}\big(\rho \mathcal{O}(\bm{\theta})\big)$ yields positive values for one class of quantum states and non-positive values for the other. This process is closely related to the concept of entanglement witnesses(as seen in Section.\ref{computable:entanglement}). This discovery shows that supervised QML could supersede entanglement witnesses when supported by high-quality quantum datasets.

\begin{figure}[H]
	\centering
	\subfigure[]{
		\label{state:space:a}
		\includegraphics[width=0.45\textwidth]{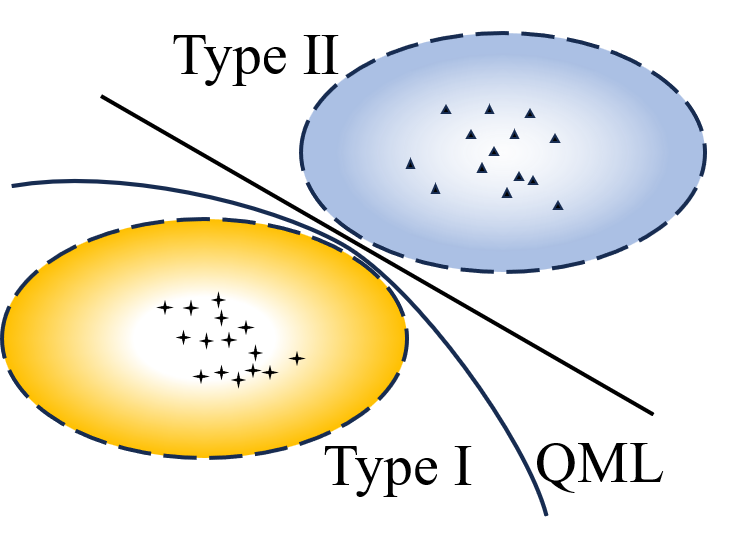}
	}
	\hspace{0.01\textwidth} 
	\subfigure[]{
		\label{state:space:b}
		\includegraphics[width=0.45\textwidth]{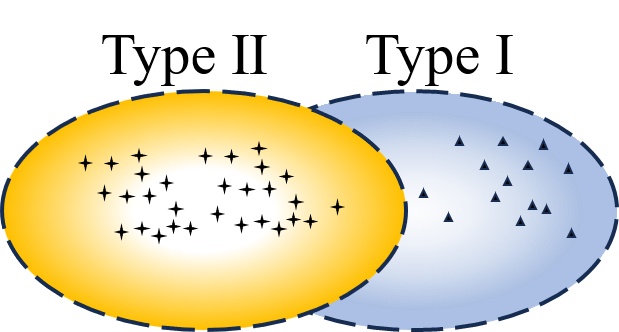}
	}
	\caption{\small \textbf{Two scenarios in supervised QML.} In the state space, we expand the state sets of different categories into convex hulls. (a) The QML model, including both linear (straight line) and nonlinear (curve) boundaries, serves as a boundary between different types of states. (b)  The two types of states exhibit inseparability by QML.}
	\label{state_space}
\end{figure}

Moreover, framing supervised QML in terms of entanglement witnesses enhances its interpretability and offers initial insights into its limitations. In Fig.~\ref{state_space}, we present two distinct scenarios: 

(a) When there exists a supporting hyperplane between the convex hulls of two state sets in the state space, supervised QML can effectively accomplish the task. The relationship between the entangled mixed-state dataset and the separable-state dataset generated in this paper is precisely such a case.

(b) When there is no supporting hyperplane between the convex hulls of two quantum state sets, supervised QML cannot perform effective classification\cite{zhang2023entanglement,lu2018separability}. For example, without the aid of other algorithmic optimizations, supervised QML is not suitable for entanglement-separability classification of the Werner states.

We believe that such binary classification problems are to be addressed, appropriate classical post-processing methods should be chosen. Even when the accuracy is high, entanglement witnesses should be combined for analysis; otherwise, the learning motivation cannot be explained. We will not further discuss such work in this paper but leave it for future research.

\section{Conclusion and discussion}\label{conclude}
In this work, we have explored the generation of entangled mixed-state datasets and their application in benchmarking QNN models on entangled-separable tasks. We have introduced a framework to generate entangled mixed states using quantum circuits, leveraging the concentratable entanglement measures and supervised quantum machine learning. To establish a theoretical foundation for the generation of entangled mixed states, we proved a continuity bound for CEL. This bound ensures that states close to each other in trace distance will have similar CEL values, providing a rigorous basis for our dataset generation approach. We demonstrated the generation of mixed states using different ansätze (HWE, SEA, and SD) and analyzed their performance in terms of generating entangled states with specific values.  To provide an interpretation of supervised QML, we have connected supervised QML with entanglement witnesses and have preliminarily outlined the limitation of supervised QML. Additionally, we conducted benchmark tests on QML models using the generated mixed-state datasets for 2, 3, 4, and 5 qubits. Our approach not only provides a valuable resource for benchmarking QML models but also opens new avenues for exploring the rich structure of quantum entanglement in mixed states. We believe that our findings will contribute to the development of more efficient and accurate methods for QML model trainin and entanglement detection, ultimately advancing the field of quantum information processing.

There are several avenues for future research remain open. First, further exploration of entanglement measures and their computational efficiency is considerable. Although the CEL provides a useful tool, more accurate and computationally feasible measures are still required to enhance the quality of the dataset. Second, the scalability of our approach to larger systems (more qubits) needs to be explored. Practical limitations, such as decoherence and gate errors, must be addressed to ensure the feasibility of generating entangled states on real quantum hardware. Finally, the generated entangled mixed-state datasets can be used in various quantum information processing tasks, such as quantum communication, quantum cryptography, quantum metrology, novel entanglement witness finding. Exploring these applications could provide new insights into the practical use of entangled states.

\bibliographystyle{unsrt}
\bibliography{clasiffy.bib}

\begin{thebibliography}{10}

\bibitem{deng2012mnist}
Li~Deng.
\newblock The mnist database of handwritten digit images for machine learning
  research [best of the web].
\newblock {\em IEEE signal processing magazine}, 29(6):141--142, 2012.

\bibitem{deng2009imagenet}
Jia Deng, Wei Dong, Richard Socher, Li-Jia Li, Kai Li, and Li~Fei-Fei.
\newblock Imagenet: A large-scale hierarchical image database.
\newblock In {\em 2009 IEEE conference on computer vision and pattern
  recognition}, pages 248--255. Ieee, 2009.

\bibitem{bennett2007netflix}
James Bennett and Stan Lanning.
\newblock The netflix prize.
\newblock 2007.

\bibitem{atienza2020advanced}
Rowel Atienza.
\newblock {\em Advanced Deep Learning with TensorFlow 2 and Keras: Apply DL,
  GANs, VAEs, deep RL, unsupervised learning, object detection and
  segmentation, and more}.
\newblock Packt Publishing Ltd, 2020.

\bibitem{krizhevsky2012imagenet}
Alex Krizhevsky, Ilya Sutskever, and Geoffrey~E Hinton.
\newblock Imagenet classification with deep convolutional neural networks.
\newblock {\em Advances in neural information processing systems}, 25, 2012.

\bibitem{cohen2017emnist}
Gregory Cohen, Saeed Afshar, Jonathan Tapson, and Andre Van~Schaik.
\newblock Emnist: Extending mnist to handwritten letters.
\newblock In {\em 2017 international joint conference on neural networks
  (IJCNN)}, pages 2921--2926. IEEE, 2017.

\bibitem{li2017cifar10}
Hongmin Li, Hanchao Liu, Xiangyang Ji, Guoqi Li, and Luping Shi.
\newblock Cifar10-dvs: an event-stream dataset for object classification.
\newblock {\em Frontiers in neuroscience}, 11:309, 2017.

\bibitem{biamonte2017quantum}
Jacob Biamonte, Peter Wittek, Nicola Pancotti, Patrick Rebentrost, Nathan
  Wiebe, and Seth Lloyd.
\newblock Quantum machine learning.
\newblock {\em Nature}, 549(7671):195--202, 2017.

\bibitem{cerezo2022challenges}
Marco Cerezo, Guillaume Verdon, Hsin-Yuan Huang, Lukasz Cincio, and Patrick~J
  Coles.
\newblock Challenges and opportunities in quantum machine learning.
\newblock {\em Nature computational science}, 2(9):567--576, 2022.

\bibitem{perrier2022qdataset}
Elija Perrier, Akram Youssry, and Chris Ferrie.
\newblock Qdataset, quantum datasets for machine learning.
\newblock {\em Scientific data}, 9(1):582, 2022.

\bibitem{park2024aqua}
Soohyun Park, Hankyul Baek, Jung~Won Yoon, Youn~Kyu Lee, and Joongheon Kim.
\newblock Aqua: Analytics-driven quantum neural network (qnn) user assistance
  for software validation.
\newblock {\em Future Generation Computer Systems}, 159:545--556, 2024.

\bibitem{alrikabi2022face}
HTS ALRikabi, Ibtisam~A Aljazaery, Jaafar~Sadiq Qateef, Abdul Hadi~M Alaidi,
  and M~Roa\textquotesingle~a.
\newblock Face patterns analysis and recognition system based on quantum neural
  network qnn.
\newblock {\em iJIM}, 16(08):35, 2022.

\bibitem{jeswal2019recent}
SK~Jeswal and S~Chakraverty.
\newblock Recent developments and applications in quantum neural network: A
  review.
\newblock {\em Archives of Computational Methods in Engineering},
  26(4):793--807, 2019.

\bibitem{sun2016quantum}
Qiming Sun and Garnet Kin-Lic Chan.
\newblock Quantum embedding theories.
\newblock {\em Accounts of chemical research}, 49(12):2705--2712, 2016.

\bibitem{lloyd2020quantum}
Seth Lloyd, Maria Schuld, Aroosa Ijaz, Josh Izaac, and Nathan Killoran.
\newblock Quantum embeddings for machine learning.
\newblock {\em arXiv preprint arXiv:2001.03622}, 2020.

\bibitem{schatzki2021entangled}
Louis Schatzki, Andrew Arrasmith, Patrick~J Coles, and Marco Cerezo.
\newblock Entangled datasets for quantum machine learning.
\newblock {\em arXiv preprint arXiv:2109.03400}, 2021.

\bibitem{kubler2021inductive}
Jonas K{\"u}bler, Simon Buchholz, and Bernhard Sch{\"o}lkopf.
\newblock The inductive bias of quantum kernels.
\newblock {\em Advances in Neural Information Processing Systems},
  34:12661--12673, 2021.

\bibitem{mandl2023linear}
Alexander Mandl, Johanna Barzen, Marvin Bechtold, Michael Keckeisen, Frank
  Leymann, and Patrick~KS Vaudrevange.
\newblock Linear structure of training samples in quantum neural network
  applications.
\newblock In {\em International Conference on Service-Oriented Computing},
  pages 150--161. Springer, 2023.

\bibitem{sharma2022trainability}
Kunal Sharma, Marco Cerezo, Lukasz Cincio, and Patrick~J Coles.
\newblock Trainability of dissipative perceptron-based quantum neural networks.
\newblock {\em Physical Review Letters}, 128(18):180505, 2022.

\bibitem{adam2019no}
Stavros~P Adam, Stamatios-Aggelos~N Alexandropoulos, Panos~M Pardalos, and
  Michael~N Vrahatis.
\newblock No free lunch theorem: A review.
\newblock {\em Approximation and optimization: Algorithms, complexity and
  applications}, pages 57--82, 2019.

\bibitem{ouyang2022training}
Long Ouyang, Jeffrey Wu, Xu~Jiang, Diogo Almeida, Carroll Wainwright, Pamela
  Mishkin, Chong Zhang, Sandhini Agarwal, Katarina Slama, Alex Ray, et~al.
\newblock Training language models to follow instructions with human feedback.
\newblock {\em Advances in neural information processing systems},
  35:27730--27744, 2022.

\bibitem{bai2022training}
Yuntao Bai, Andy Jones, Kamal Ndousse, Amanda Askell, Anna Chen, Nova DasSarma,
  Dawn Drain, Stanislav Fort, Deep Ganguli, Tom Henighan, et~al.
\newblock Training a helpful and harmless assistant with reinforcement learning
  from human feedback.
\newblock {\em arXiv preprint arXiv:2204.05862}, 2022.

\bibitem{wang2024transition}
Xinbiao Wang, Yuxuan Du, Zhuozhuo Tu, Yong Luo, Xiao Yuan, and Dacheng Tao.
\newblock Transition role of entangled data in quantum machine learning.
\newblock {\em Nature Communications}, 15(1):3716, 2024.

\bibitem{sharma2022reformulation}
Kunal Sharma, Marco Cerezo, Zo{\"e} Holmes, Lukasz Cincio, Andrew Sornborger,
  and Patrick~J Coles.
\newblock Reformulation of the no-free-lunch theorem for entangled datasets.
\newblock {\em Physical Review Letters}, 128(7):070501, 2022.

\bibitem{nakayama2023vqe}
Akimoto Nakayama, Kosuke Mitarai, Leonardo Placidi, Takanori Sugimoto, and
  Keisuke Fujii.
\newblock Vqe-generated quantum circuit dataset for machine learning.
\newblock {\em arXiv preprint arXiv:2302.09751}, 2023.

\bibitem{placidi2023mnisq}
Leonardo Placidi, Ryuichiro Hataya, Toshio Mori, Koki Aoyama, Hayata Morisaki,
  Kosuke Mitarai, and Keisuke Fujii.
\newblock Mnisq: A large-scale quantum circuit dataset for machine learning
  on/for quantum computers in the nisq era.
\newblock {\em arXiv preprint arXiv:2306.16627}, 2023.

\bibitem{horodecki2009quantum}
Ryszard Horodecki, Pawe{\l} Horodecki, Micha{\l} Horodecki, and Karol
  Horodecki.
\newblock Quantum entanglement.
\newblock {\em Reviews of modern physics}, 81(2):865--942, 2009.

\bibitem{branciard2013measurement}
Cyril Branciard, Denis Rosset, Yeong-Cherng Liang, and Nicolas Gisin.
\newblock Measurement-device-independent entanglement witnesses for all
  entangled quantum states.
\newblock {\em Physical review letters}, 110(6):060405, 2013.

\bibitem{sperling2013multipartite}
Jan Sperling and Werner Vogel.
\newblock Multipartite entanglement witnesses.
\newblock {\em Physical review letters}, 111(11):110503, 2013.

\bibitem{chruscinski2014entanglement}
Dariusz Chru{\'s}ci{\'n}ski and Gniewomir Sarbicki.
\newblock Entanglement witnesses: construction, analysis and classification.
\newblock {\em Journal of Physics A: Mathematical and Theoretical},
  47(48):483001, 2014.

\bibitem{cao2024genuine}
Huan Cao, Simon Morelli, Lee~A Rozema, Chao Zhang, Armin Tavakoli, and Philip
  Walther.
\newblock Genuine multipartite entanglement detection with imperfect
  measurements: Concept and experiment.
\newblock {\em Physical Review Letters}, 133(15):150201, 2024.

\bibitem{PhysRevLett.127.140501}
Jacob~L. Beckey, N.~Gigena, Patrick~J. Coles, and M.~Cerezo.
\newblock Computable and operationally meaningful multipartite entanglement
  measures.
\newblock {\em Phys. Rev. Lett.}, 127:140501, Sep 2021.

\bibitem{walborn2007experimental}
SP~Walborn, PH~Souto Ribeiro, L~Davidovich, F~Mintert, and A~Buchleitner.
\newblock Experimental determination of entanglement by a projective
  measurement.
\newblock {\em Physical Review A-Atomic, Molecular, and Optical Physics},
  75(3):032338, 2007.

\bibitem{walborn2006experimental}
SP~Walborn, PH~Souto~Ribeiro, L~Davidovich, F~Mintert, and A~Buchleitner.
\newblock Experimental determination of entanglement with a single measurement.
\newblock {\em Nature}, 440(7087):1022--1024, 2006.

\bibitem{uhlmann2010roofs}
Armin Uhlmann.
\newblock Roofs and convexity.
\newblock {\em Entropy}, 12(7):1799--1832, 2010.

\bibitem{beckey2023multipartite}
Jacob~L Beckey, Gerard Pelegr{\'\i}, Steph Foulds, and Natalie~J Pearson.
\newblock Multipartite entanglement measures via bell-basis measurements.
\newblock {\em Physical Review A}, 107(6):062425, 2023.

\bibitem{Foulds_2024}
Steph Foulds, Oliver Prove, and Viv Kendon.
\newblock Generalizing multipartite concentratable entanglement for practical
  applications: mixed, qudit and optical states.
\newblock {\em Philosophical Transactions of the Royal Society A: Mathematical,
  Physical and Engineering Sciences}, 382(2287), December 2024.

\bibitem{schuld2015introduction}
Maria Schuld, Ilya Sinayskiy, and Francesco Petruccione.
\newblock An introduction to quantum machine learning.
\newblock {\em Contemporary Physics}, 56(2):172--185, 2015.

\bibitem{zhang2020recent}
Yao Zhang and Qiang Ni.
\newblock Recent advances in quantum machine learning.
\newblock {\em Quantum Engineering}, 2(1):e34, 2020.

\bibitem{batra2021quantum}
Kushal Batra, Kimberley~M Zorn, Daniel~H Foil, Eni Minerali, Victor~O
  Gawriljuk, Thomas~R Lane, and Sean Ekins.
\newblock Quantum machine learning algorithms for drug discovery applications.
\newblock {\em Journal of chemical information and modeling}, 61(6):2641--2647,
  2021.

\bibitem{lloyd2013quantum}
Seth Lloyd, Masoud Mohseni, and Patrick Rebentrost.
\newblock Quantum algorithms for supervised and unsupervised machine learning.
\newblock {\em arXiv preprint arXiv:1307.0411}, 2013.

\bibitem{schuld2021supervised}
Maria Schuld.
\newblock Supervised quantum machine learning models are kernel methods.
\newblock {\em arXiv preprint arXiv:2101.11020}, 2021.

\bibitem{alvarez2017supervised}
Unai Alvarez-Rodriguez, Lucas Lamata, Pablo Escandell-Montero, Jos{\'e}~D
  Mart{\'\i}n-Guerrero, and Enrique Solano.
\newblock Supervised quantum learning without measurements.
\newblock {\em Scientific reports}, 7(1):13645, 2017.

\bibitem{zhang2023entanglement}
Lifeng Zhang, Zhihua Chen, and Shao-Ming Fei.
\newblock Entanglement verification with deep semisupervised machine learning.
\newblock {\em Physical Review A}, 108(2):022427, 2023.

\bibitem{lu2018separability}
Sirui Lu, Shilin Huang, Keren Li, Jun Li, Jianxin Chen, Dawei Lu, Zhengfeng Ji,
  Yi~Shen, Duanlu Zhou, and Bei Zeng.
\newblock Separability-entanglement classifier via machine learning.
\newblock {\em Physical Review A}, 98(1):012315, 2018.

\bibitem{dur2000three}
Wolfgang D{\"u}r, Guifre Vidal, and J~Ignacio Cirac.
\newblock Three qubits can be entangled in two inequivalent ways.
\newblock {\em Physical Review A}, 62(6):062314, 2000.

\bibitem{pittenger2000note}
Arthur~O Pittenger and Morton~H Rubin.
\newblock Note on separability of the werner states in arbitrary dimensions.
\newblock {\em Optics Communications}, 179(1-6):447--449, 2000.

\bibitem{gao2010detection}
Ting Gao and Yan Hong.
\newblock Detection of genuinely entangled and nonseparable n-partite quantum
  states.
\newblock {\em Physical Review A-Atomic, Molecular, and Optical Physics},
  82(6):062113, 2010.

\bibitem{Nielsen_Chuang_2010}
Michael~A. Nielsen and Isaac~L. Chuang.
\newblock {\em Quantum Computation and Quantum Information: 10th Anniversary
  Edition}.
\newblock Cambridge University Press, 2010.

\bibitem{riedel2021bell}
Elias Riedel~G{\aa}rding, Nicolas Schwaller, Chun~Lam Chan, Su~Yeon Chang,
  Samuel Bosch, Frederic Gessler, Willy~Robert Laborde, Javier~Naya Hernandez,
  Xinyu Si, Marc-Andr{\'e} Dupertuis, et~al.
\newblock Bell diagonal and werner state generation: Entanglement,
  non-locality, steering and discord on the ibm quantum computer.
\newblock {\em Entropy}, 23(7):797, 2021.

\bibitem{cruz2019efficient}
Diogo Cruz, Romain Fournier, Fabien Gremion, Alix Jeannerot, Kenichi Komagata,
  Tara Tosic, Jarla Thiesbrummel, Chun~Lam Chan, Nicolas Macris, Marc-Andr{\'e}
  Dupertuis, et~al.
\newblock Efficient quantum algorithms for ghz and w states, and implementation
  on the ibm quantum computer.
\newblock {\em Advanced Quantum Technologies}, 2(5-6):1900015, 2019.

\bibitem{kandala2017hardware}
Abhinav Kandala, Antonio Mezzacapo, Kristan Temme, Maika Takita, Markus Brink,
  Jerry~M Chow, and Jay~M Gambetta.
\newblock Hardware-efficient variational quantum eigensolver for small
  molecules and quantum magnets.
\newblock {\em nature}, 549(7671):242--246, 2017.

\bibitem{schuld2020circuit}
Maria Schuld, Alex Bocharov, Krysta~M Svore, and Nathan Wiebe.
\newblock Circuit-centric quantum classifiers.
\newblock {\em Physical Review A}, 101(3):032308, 2020.

\bibitem{cerezo2021cost}
Marco Cerezo, Akira Sone, Tyler Volkoff, Lukasz Cincio, and Patrick~J Coles.
\newblock Cost function dependent barren plateaus in shallow parametrized
  quantum circuits.
\newblock {\em Nature communications}, 12(1):1791, 2021.

\bibitem{vakili2020performance}
Meysam Vakili, Mohammad Ghamsari, and Masoumeh Rezaei.
\newblock Performance analysis and comparison of machine and deep learning
  algorithms for iot data classification.
\newblock {\em arXiv preprint arXiv:2001.09636}, 2020.

\bibitem{sokolova2006beyond}
Marina Sokolova, Nathalie Japkowicz, and Stan Szpakowicz.
\newblock Beyond accuracy, f-score and roc: a family of discriminant measures
  for performance evaluation.
\newblock In {\em Australasian joint conference on artificial intelligence},
  pages 1015--1021. Springer, 2006.

\end{thebibliography}
\appendix
\numberwithin{equation}{section}
\begin{appendices}
	\setcounter{equation}{0}
	\section{Detailed Calculations}\label{app:details}
	\textbf{GHZ state with white noise:} We begin with a trivial subset $\bar{\alpha}$, where $\bar{\alpha}\in \mathcal{P}(S)$. Denote the number of elements in $\bar{\alpha}$ as $|\bar{\alpha}|$, and let $|\bar{\alpha}|=j \geq 1$. The reduced state yielded by tracing over the qubits in $\bar{\alpha}$ can be written as
	\begin{equation}
		\begin{split}
			\rho_{\alpha}=Tr_{\bar{\alpha}}[\rho]&=\frac{p}{2}(\ket{0}\bra{0}^{\otimes n -j}+\ket{1}\bra{1}^{\otimes n -j})+\frac{1-p}{2^n}\mathcal{I}^{\otimes n -j}.
		\end{split}
	\end{equation}
	So the square of the reduced state is
	\begin{equation}
		\begin{split}
			\rho_{\alpha}^2 &=\frac{p^2}{4}(\ket{0}\bra{0}^{\otimes n -j}+\ket{1}\bra{1}^{\otimes n -j})+\frac{(1-p)^2}{4^{n-j}}\mathcal{I}^{\otimes n -j}\\
			&+\frac{p(1-p)}{2^{n-j}}(\ket{0}\bra{0}^{\otimes n -j}+\ket{1}\bra{1}^{\otimes n -j}) \\
		\end{split}
	\end{equation}
	Next, for any $\alpha \in \mathcal{P}(S)$, we have
	\begin{equation}
		tr[\rho_{\alpha}^2]=\frac{p^2}{2}+\frac{(1-p)^2}{2^{n-j}}+\frac{p(1-p)}{2^{n-j-1}}.
	\end{equation}

	Applying the lower bound of CE, we get
	\begin{equation}
		\begin{split}
			C^{l}_{\rho}(S)&=\frac{1}{2^n}+(1-\frac{1}{2^n})tr[\rho^2]-\frac{1}{2^n}(1-tr[\rho^2])-\frac{1}{2^n}\sum_{\alpha \in \mathcal{\Tilde{P}}(S)} tr[\rho_{\alpha}^2]\\
			&=(1-\frac{1}{2^{n-1}})tr[\rho^2]-\frac{1}{2^n}\sum_{j=1}^{n-1} \binom{n}{j}(\frac{p^2}{2}+\frac{(1-p)^2}{2^{n-j}}+\frac{p(1-p)}{2^{n-j-1}})\\
			&=(1-\frac{1}{2^{n-1}})tr[\rho^2]-\frac{(2^{n-1}-1)p^2}{2^n}-\frac{1-p^2}{4^n}(3^n-2^n-1)\\
			&=\frac{p^2}{2} - \frac{1}{2^n} + (1 - p^2) \left( \frac{3}{2^n} - \frac{1}{4^n} - \left( \frac{3}{4} \right)^n \right)
		\end{split}
	\end{equation}
	
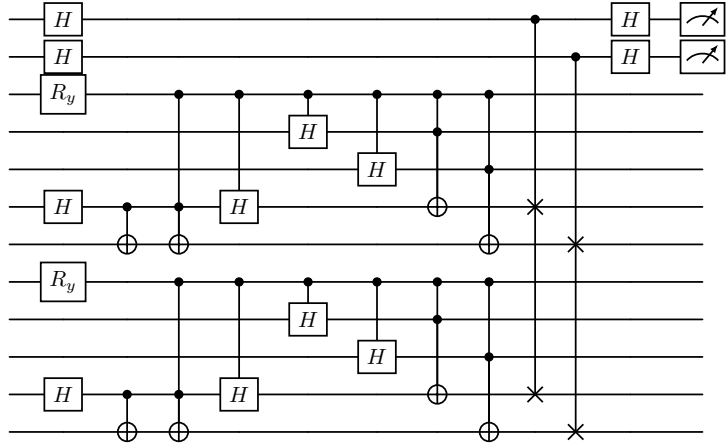
\begin{figure}[H]
		\centering
		\resizebox{0.8\textwidth}{!}{%
			\begin{quantikz}[row sep={0.6cm,between origins},scale=0.7]
				&\gate{H}&&&&&&&& \ctrl{5} &&\gate{H}&\meter{}\\
				&\gate{H}&&&&&&&&&\ctrl{5}&\gate{H}&\meter{}\\
				& \gate{R_y} && \ctrl{4}&\ctrl{3}&\ctrl{1}&\ctrl{2}&\ctrl{3}&\ctrl{4}& &&&\\
				&&&&&\gate{H}&&\ctrl{2}&&&&&\\
				&&&&&&\gate{H}&&\ctrl{2}&&&&\\ 
				&\gate{H}& \ctrl{1}&\ctrl{1}&\gate{H}&&&\targ{}&& \swap{5}&&& \\ 
				&&\targ{}&\targ{}&&&&&\targ{}&&\swap{5}&&\\
				& \gate{R_y} && \ctrl{4}&\ctrl{3}&\ctrl{1}&\ctrl{2}&\ctrl{3}&\ctrl{4}&&& &\\
				&&&&&\gate{H}&&\ctrl{2}&&&&&\\
				&&&&&&\gate{H}&&\ctrl{2}&&&&\\ 
				&\gate{H}& \ctrl{1}&\ctrl{1}&\gate{H}&&&\targ{}&&\targX{} &&&\\ 
				&&\targ{}&\targ{}&&&&&\targ{}&&\targX{}&&
			\end{quantikz}%
		}
		\caption{\small \textbf{Quantum Circuit for GHZ state with white noise.} Here, we present the parallel swap test quantum circuit for a 2-qubit W state with white noise. The probability $p$ mentioned above is controlled by the $R_y(\phi)$ gate in the circuit, where $p=cos^{2}(\frac{\phi}{2})$}
		\label{CELGHZ}
	\end{figure}

	In Fig.~\ref{CELGHZ}, we provide the detailed quantum circuit for computing the CEL of a 2-qubit GHZ state with white noise. For the \(n\)-qubit case, \(5n + 2\) qubits, including \(n\) auxiliary qubits, are required.
	
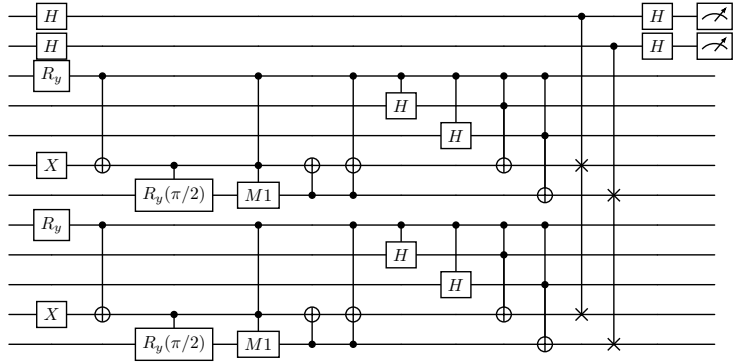
\begin{figure}[h!]
	     \centering
	     \resizebox{0.8\textwidth}{!}{%
		     \begin{quantikz}[row sep={0.6cm,between origins},scale=0.7]
		     	  &\gate{H}&&&&&&&&&&\ctrl{5}&&\gate{H}&\meter{}\\
		     	  &\gate{H}&&&&&&&&&&&\ctrl{5}&\gate{H}&\meter{}\\
			       &\gate{R_y}&\ctrl{3}&                 &\ctrl{3} &         &\ctrl{3}   &\ctrl{1}&\ctrl{2}&\ctrl{3}&\ctrl{4}&&&&\\
			      &          &		   &                 &         &         &         &\gate{H}&        &\ctrl{2}&&&&&\\
			      &          &		   &                 &         &         &         &        &\gate{H}&        &\ctrl{2}&&&&\\
			      &\gate{X}  &\targ{} &\ctrl{1}         &\ctrl{1} &\targ{}  &\targ{}  &        &        &\targ{} &&\swap{5}&&&\\
			      &		  &		   &\gate{R_y(\pi/2)}&\gate{M1}&\ctrl{-1}&\ctrl{-1}&        &        &        &\targ{}&&\swap{5}&&\\
			      &\gate{R_y}&\ctrl{3}&                 &\ctrl{3} &         &\ctrl{3}   &\ctrl{1}&\ctrl{2}&\ctrl{3}&\ctrl{4}&&&&\\
			      &          &		   &                 &         &         &         &\gate{H}&        &\ctrl{2}&&&&&\\
			      &          &		   &                 &         &         &         &        &\gate{H}&        &\ctrl{2}&&&&\\
			     &\gate{X}  &\targ{} &\ctrl{1}         &\ctrl{1} &\targ{}  &\targ{}  &        &        &\targ{} &&\targX{}&&&\\
			      &		  &		   &\gate{R_y(\pi/2)}&\gate{M1}&\ctrl{-1}&\ctrl{-1}&        &        &        &\targ{}&&\targX{}&&\\
			     \end{quantikz}%
		     }
	  \caption{\small \textbf{Quantum Circuit for W state with white noise.} 
	  	Here, we present the parallel swap test quantum circuit for a 2-qubit W state with white noise. The circuit design for the W state with white noise is referenced from \cite{cruz2019efficient}. For the 2-qubit case, the matrix $M1$ is given by:
	  	$ M1 = \begin{pmatrix} \sqrt{1/2} & -\sqrt{1/2} \\ \sqrt{1/2} & \sqrt{1/2} \end{pmatrix} $.
	  }
	     \label{CElWcircuit}
	 \end{figure}

	\textbf{W state with white noise:} For the W state, we need a recursive formula in the following:
	\begin{equation}
		\ket{W_2}=\frac{1}{\sqrt{2}}({\ket{10}+\ket{01}})
	\end{equation}
	\begin{equation}\label{recursive}
		\ket{W_n}=\frac{\sqrt{n-1}}{\sqrt{n}}\ket{W_{n-1}}\otimes\ket{0}+\frac{1}{\sqrt{n}}\ket{0_{n-1}}\otimes\ket{1}
	\end{equation}
	This partition is not unique; in other words, we can assume that the systems reduced successively all appear in the last subsystem as \eqref{recursive}. We begin with a trivial subset $\bar{\alpha}$, where $\bar{\alpha}\in \mathcal{P}(S)$. Denote the number of elements in $\bar{\alpha}$ as $|\bar{\alpha}|$, and let $|\bar{\alpha}|=j \geq 1$. The reduced state yielded by tracing over the qubits in $\bar{\alpha}$ can be written as
	
	\begin{equation}
		\begin{split}
			\rho_{\alpha}&=Tr_{\bar{\alpha}}(\rho)\\
			&=Tr_{\bar{\alpha}}[p(\frac{n-1}{n}\ket{W_{n-1}}\bra{W_{n-1}}\otimes\ket{0}\bra{0}+\frac{1}{n}\ket{0_{n-1}}\bra{0_{n-1}}\otimes\ket{1}\bra{1}\\
			&+\frac{\sqrt{n-1}}{\sqrt{n}}\ket{W_{n-1}}\bra{0_{n-1}}\otimes\ket{0}\bra{1}+\frac{\sqrt{n-1}}{\sqrt{n}}\ket{0_{n-1}}\bra{W_{n-1}}\otimes\ket{1}\bra{0})\\
			&+\frac{1-p}{2^n}I^{\otimes n}]\\
			&=p(\frac{n-j}{n}\ket{W_{n-j}}\bra{W_{n-j}}+\frac{j}{n}\ket{0_{n-j}}\bra{0_{n-j}})+\frac{1-p}{2^{n-j}}I^{\otimes n-j}\\
		\end{split}
	\end{equation}
	Then we can have
	\begin{equation}
		\begin{split}
			tr[\rho_{\alpha}^2]&=p^2(\frac{(n-j)^2}{n^2}+\frac{j^2}{n^2})+p(\frac{(1-p)^2}{2^{n-j}}+\frac{1-p}{2^{n-1-j}})
		\end{split}
	\end{equation}
	So the square of the reduced state can be finally written as
	\begin{equation}
		\begin{split}
			C^{l}_{\rho}&=\frac{1}{2^n}+(1-\frac{1}{2^n}tr[\rho^2])-\frac{1}{2^n}\sum_{\alpha \in \mathcal{P}(S)} tr[\rho_{\alpha}^2]\\
			&=\frac{1}{2^n}+(1-\frac{1}{2^n}tr[\rho^2])-\frac{1}{2^n}(1+tr[\rho^2])-\frac{1}{2^n}\sum_{\alpha \in \bar{\mathcal{P}}(S)} tr[\rho_{\alpha}^2]\\
			&=(1-\frac{1}{2^{n-1}})(\frac{1-p^2}{2^n}+p^2)-\frac{p^2}{2^n}(\frac{2^{n-1}(n+1)}{n}-2)\\
			&-\frac{(1-p^2)(3^n-2^n-1)}{4^n}\\
			&=\frac{(n-1)p^2}{2n} + (1 - p^2) \left( \frac{2}{2^n} - \frac{1}{4^n} - \left( \frac{3}{4} \right)^n \right)
		\end{split}
	\end{equation}
In Fig.~\ref{CElWcircuit}, we provide the detailed quantum circuit for computing the CEL of a 2-qubit W state with white noise. For the \(n\)-qubit case, \(5n + 2\) qubits, including \(n\) auxiliary qubits, are required.

\end{appendices}
\newpage
\begin{appendices}
	\setcounter{equation}{0}
	\section{Proof of Thereom 1}\label{app:Thm}
	In this section, we provide a proof for Theorem\ref{thm} in the main text. In Ref.~\cite{schatzki2021entangled}, the authors showed that for any reduced density matrix for index subset $\alpha \in \mathcal{P}(S)$, for given $\rho$ and $\sigma$, $|Tr(\rho_{\alpha}^2-\sigma_{\alpha}^2)| \leq \sqrt{2}||\rho-\sigma||_1$. One can easily obtain :
	\begin{equation}
		\begin{split}
			|C^{l}_{\rho}(S)-C^{l}_{\sigma}(S)| &= \left|\frac{1}{2^n} + \left(1 - \frac{1}{2^n}\operatorname{tr}[\rho^2]\right) - \frac{1}{2^n}\sum_{\alpha \in \mathcal{P}(S)} \operatorname{tr}[\rho_{\alpha}^2] \right. \\
			&\quad \left. - \frac{1}{2^n} + \left(1 - \frac{1}{2^n}\operatorname{tr}[\sigma^2]\right) - \frac{1}{2^n}\sum_{\alpha \in \mathcal{P}(S)} \operatorname{tr}[\sigma_{\alpha}^2]\right| \\
			&= \frac{1}{2^n}\left|\operatorname{tr}[\sigma^2] - \operatorname{tr}[\rho^2] + \sum_{\alpha \in \mathcal{P}(S)} \left(\operatorname{tr}[\sigma_{\alpha}^2] - \operatorname{tr}[\rho_{\alpha}^2]\right)\right| \\
			&\leq \frac{1}{2^n}\left(\left|\operatorname{tr}[\sigma^2] - \operatorname{tr}[\rho^2]\right| + \sum_{\alpha \in \mathcal{P}(S)} \left|\operatorname{tr}[\rho_{\alpha}^2] - \operatorname{tr}[\sigma_{\alpha}^2]\right|\right) \\
			&\leq \left(\frac{1}{2^n} + 1\right)\sqrt{2}\|\rho - \sigma\|_1 \\
			&\leq \left(\frac{1}{2^n} + 1\right)\sqrt{2}d
		\end{split}
	\end{equation}
\end{appendices}

\end{document}